\documentclass{article}

\usepackage{graphicx}

\def\ligne#1{\hbox to \hsize{#1}}
\def\PlacerEn#1 #2 #3 {\rlap{\kern#1\raise#2\hbox{#3}}}

\newtheorem{fig}{Figure}

\newtheorem{thm}{Theorem}

\newtheorem{tab}{Table}

\def\leurre{\noindent\leftskip0pt\small\baselineskip 10pt}
\def\grostrait{\ligne{\vrule height 1pt depth 1pt width \hsize}}
\def\demitrait{\ligne{\vrule height 0.5pt depth 0.5pt width \hsize}}
\def\ttV{\vrule height 12pt depth 6pt width 0pt}

\setbox221=\hbox{\includegraphics{cercle_L20.eps}}

\def\encercle#1#2{\hbox{\raise-5pt\copy221\hskip#2#1}}

\title{An application of Grossone to the study of a family of tilings
of the hyperbolic plane}
\author{Maurice Margenstern\\
Laboratoire d'Informatique Th\'eorique et Appliqu\'ee, EA 3097,\\
        Universit\'e Paul Verlaine $-$ Metz, UFR-MIM, \\
        \^Ile du Saulcy, 57045 Metz Cedex, France\\
and CNRS, LORIA\\
{\it e-mail}: {\tt margens@univ-metz.fr}
}

\begin{document}
\maketitle

\begin{abstract}
   In this paper, we look at the improvement of our knowledge on
a family of tilings of the hyperbolic plane which is brought in by the use of
Sergeyev's numeral system based on grossone, see~\cite{sergeyev1,sergeyev2,sergeyev3}. 
It appears that the information we can get
by using this new numeral system depends on the way we look at the tilings. The ways are 
significantly different but they confirm some results which were obtained in the 
traditional but constructive frame and allow us to obtain an additional precision with 
respect to this information. 
\end{abstract}

\def\cqfd{\hbox{\kern 2pt\vrule height 6pt depth 2pt width 8pt\kern 1pt}}
\def\Hii{$I\!\!H^2$}
\def\Hiii{$I\!\!H^3$}
\def\Hiv{$I\!\!H^4$}
\def\norm{\hbox{$\vert\vert$}}
\def\grossone{\hbox{\textcircled{\bf 1}}}
\section{{\Large Introduction}}

   This paper gives an application of the new methodology introduced by Yaroslav 
{\sc Sergeyev} in his seminal papers, see~\cite{sergeyev1,sergeyev2,sergeyev3},
to the study of a family of tilings of the hyperbolic plane.

   The hyperbolic plane is of great interest by itself: both for theoretical reasons, 
as Section~\ref{hypgeom} will convince the reader and also for practical ones.
Up to now, the main applications of hyperbolic geometry are theoretical and they concern
the theory of relativity. Recently, a few applications were planned and a few of them
realized, we refer the interested reader to~\cite{mmbook2,mmHongKong}. From
what is explained in~\cite{mmbook2,mmHongKong}, we can infer that hyperbolic geometry,
and especially the location technique of tiles described in~\cite{mmbook1}
and in Section~\ref{hyptilings}, can be of use for issues involving huge nets. 

   In Section~\ref{grossone}, we remind the basic features of the new numeral system 
which allows to deal with infinite sets. Then, in Section~\ref{hypgeom} we remind what 
is needed of hyperbolic geometry in order to introduce the family of tilings which
we consider in Section~\ref{hyptilings}. In Section~\ref{hypgrossone}, we state the
results and prove them. The results presented in this paper are slightly different
from those briefly presented at the International Workshop {\it Infinite and 
Infinitesimal in Mathematics, Computing and Natural Sciences} held at Cetraro, Italy, 
in May 2010, see~\cite{mmIW}. They are in some sense a more precise version of what
was given in~\cite{mmIW}. In Section~\ref{conclusion} we indicate a few possible 
continuations.

\section{{\Large The new numeral system}}
\label{grossone}

   In papers~\cite{sergeyev1,sergeyev2,sergeyev3}, Yaroslav {\sc Sergeyev} gives the
main arguments in favour of the new numeral system he founded, allowing to obtain
more precise results on infinite sets that what was obtained previously.

   We can sum up the properties of the system as follows. 

   We distinguish the {\bf objects} of our study from the {\bf tools} we use 
to {\bf observe} them. These three parts of the knowledge process have to be more 
clearly distinguished as they were traditionally in mathematics, contrarily to other 
domains of science, as physics and natural sciences where this distinction is clearly 
observed. This is the content of {\it Postulate~$2$} in the quoted papers. It is an
important issue for mathematics where the distinction between an {\bf observer}
and what is {\bf observed} is very often forgotten. In particular, not enough
attention is paid to subjectivity of the observer and the relative validity of 
his/her observations. The latter are very dependent of cultural elements, especially 
the {\it language} used by the observer to describe what he/she sees.

   We are interested in the properties of the objects, some of them being possibly
infinite or infinitesimal, but {\bf operations} on the objects, performed by a human
being or a machine, necessarily deal with {\bf finitely many} of them and only
{\bf finitely many} operations can be applied within the frame of an argument.
This is the content of {\it Postulate~$1$} in the quoted papers.

   At last and not the least, we consider that the principle {\it The part is less than
the whole} has to be applied to all numbers, finite, infinite or infinitesimal, and
also to all sets and processes, whether finite or infinite. This is the content
of {\it Postulate~$3$} of~\cite{sergeyev1,sergeyev2,sergeyev3}.

   On the basis of these principles, Yaroslav {\sc Sergeyev} introduced a new numeral
system in order to be able to write down infinite numbers. To this aim, an {\bf infinite
natural number} is introduced, {\bf grossone}, denote by \grossone, which is the
number of elements of the set of positive integers. This number satisfies the 
following three properties which are axioms of the system:
\vskip 10pt
   {\leftskip 18pt\parindent 0pt\rightskip 18pt
   $-$ for any finite natural number~$n$, $n<\ $\grossone.

   $-$ we have \hbox{0.\grossone = \grossone.0 = 0}, 
\hbox{\grossone~$-$~\grossone = 0}, \hbox{$\displaystyle{\grossone\over\grossone}=1$},
\hbox{$\grossone^0=1$}, \hbox{$1^{\grossone}=1$} and \hbox{$0^{\grossone}=0$.}

   $-$ let $I\!\!N_{k,n}$ be the set of positive integers of the form $k$+$jn$,
with $k$ and $n$ positive finite integers, $k<n$, for $j$~running over the set of the 
positive integers; notice that these sets are pairwise disjoint and that their union 
is the set of all positive integers; then all these sets have the same number of 
elements denoted by $\displaystyle{\grossone\over n}$.
   \par}
\vskip 10pt
   Denote by $I\!\!N$ the set of positive natural numbers. All traditional operations 
performed on natural numbers are extended to~\grossone{} in a natural way with the 
standard properties, among them: commutativity and associativity of
addition and multiplication and distributivity of multiplication over addition. As
$n$ is the number of elements of the set of the finite positive integers from~1 to~$n$,
and as \grossone{} is, by definition, the number of elements of~$I\!\!N$,
a consequence of the properties of addition and multiplication, is that $I\!\!N$ also 
contains a lot of other infinite numbers: all of them of the form
$\displaystyle{\grossone\over n}$ and, more generally,
all the numbers $\displaystyle{{j\grossone}\over n}$$\pm$$k$ for any $j\in 1..n$ 
and any finite natural number~$k$, $n$ being any positive integer. 
From now on, we shall call {\bf infinite numeral system} the system described as above.

   Before turning to hyperbolic geometry and our application of this system to them,
we conclude this short introduction to the infinite numeral system by two points
about infinite numbers.

   First, let us remark that there are other infinite numbers as those described in the
previous paragraphs. Let us remark that we can define numbers by defining their 
properties. We know that any finite positive number~$n$ is the greatest element 
of the set of positive numbers~$m$ such that $m\leq n$. This definition can in fact
be extended to~\grossone{} itself which is the number of positive numbers, so that
\grossone{} itself is a number and from what is just said, it is the greatest of them.
As an other example, consider the set~$S$ of positive integers~$x$ such that 
$x^2 \leq \grossone$. The number~$\kappa$ of elements of~$S$ is also
the greatest element of~$S$, by analogy with what we have seen with any finite~$n$ 
and with~\grossone{} itself. Now, $\kappa$~is infinite. Otherwise, $\kappa$~being 
finite would entail that $\kappa$+1 would also be finite and so, $(\kappa$$+$$1)^2$ 
would also be finite. Accordingly, $\kappa$+1 would belong to~$S$,
a contradiction with the maximality of~$\kappa$. From this, we obtain
that $\grossone < (\kappa$$+$$1)^2$, so that we can write that
$\kappa=\lfloor\sqrt{\grossone}\rfloor$.  
We shall go back to this way of 
defining numbers in Section~\ref{hypgrossone}.

   Second, we shall also use {\bf sequences} of numbers. A {\bf sequence} of elements
of a set~$A$ is a mapping from the set of positive integers into a set~$A$. As a 
consequence of the above axioms, the number of elements of a sequence is at 
most \grossone. We say that a sequence is {\bf complete} if it exactly has
\grossone{} elements. 

\section{{\Large Hyperbolic geometry}}
\label{hypgeom}

Hyperbolic geometry appeared in the first half of the
19$^{\rm th}$ century, proving the independence of the parallel
axiom of Euclidean geometry. Models were devised in the second
half of the 19$^{\rm th}$ century and we shall use here one of the
most popular ones, Poincar\'e's disc. This model is represented by
Figure~\ref{poincare}. 

In the figure, the model works as follows: inside the open disc, we have the points 
of the hyperbolic plane. Lines are trace of diameters or circles orthogonal to the 
border of the disc. As an example, the line~$m$ of Figure~\ref{poincare} is such a line. 
Through the point~$A$ we can see a line~$s$ which cuts~$m$, two lines
which are parallel to~$m$: $p$ and~$q$, touching~$m$ in the model at~$P$ and~$Q$
respectively. The points $P$ and~$Q$ are points of the border. They do not belong
to the hyperbolic plane and, for this reason, they are called points at infinity.
At last, and not the least: the line~$n$ also passes through~$A$ without cutting~$m$,
neither inside the disc nor outside it. This line is called {\bf non-secant} with~$m$.
Two lines are non-secant if and only if they have a common perpendicular which is
unique.

\vskip 20pt
\setbox110=\hbox{\includegraphics[scale=1]{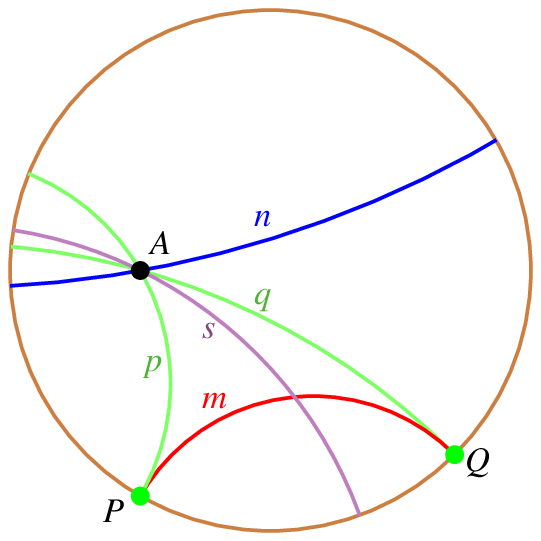}}
\vtop{
\ligne{\hfill
\PlacerEn {-285pt} {0pt} \box110
}
\vspace{-15pt}
\begin{fig} \label{poincare}
\leurre
Illustration of the parallel axiom of hyperbolic geometry in Poincar\'e's disc
model.

\end{fig}
}

   The model can be generalized to any dimension, but as we deal with the plane only in
this paper, we simply refer the reader to~\cite{mmbook1,mmbook2} where dimensions~3 and~4
are studied for further indications. More classical approaches can be found 
in~\cite{meschkowski}, \cite{millman} and~\cite{ramsay}, for instance. 
Hyperbolic geometry is used in the theory of relativity, see~\cite{varicak,ungar} and 
in several cosmological models, see~\cite{friedman}. We refer the interested reader 
to the corresponding sections of {\it Wikipedia} for more references. 
In the first subsection of the next section, we indicate more specific applications of 
tilings of the hyperbolic plane.

\section{{\Large The tilings $\{p,q\}$}}
\label{hyptilings}

   We remind the reader that a {\bf tiling} is a partition of a geometrical space~$X$
where the closures of the elements of the partition are supposed to be obtained from a 
set~$\cal S$ of parts of~$X$ by isometries of the space. We say that $\cal S$ is the 
set of {\bf prototiles}. The closures of the elements of the partition are said 
{\bf copies} of the prototiles and they are called {\bf tiles}. Moreover, there can be 
an additional condition on the abutting tiles to be satisfied: they are
called the {\bf matching} conditions.

   In this paper, we shall focus on the case where we have finitely many prototiles
which are all copies of the same polygon~$P$. Such a tiling is called a 
{\bf tessellation} when it is generated by reflection in the sides of~$P$
and, recursively, of the images in their sides. Below, Figure~\ref{tilings} 
illustrates two particular cases of tessellations to which we turn a bit later.

\subsection{Poincar\'e's theorem}

   First, we mention an important theorem proved by Poincar\'e which says that there
are infinitely many different polygons giving rise to a tessellation of the hyperbolic
plane.

\begin{thm} {\rm (Poincar\'e)} $-$
A triangle of the hyperbolic plane whose angles are of the form
\hbox{$\displaystyle{{2\pi}\over p}$},\hbox{$\displaystyle{{2\pi}\over q}$}
and \hbox{$\displaystyle{{2\pi}\over r}$}, where $p$, $q$ and $r$ are
positive integers, generates a tiling of the hyperbolic
plane by tessellation when $p$, $q$ and~$r$ satisfy the condition 
\hbox{$\displaystyle{1\over p}+\displaystyle{1\over q}+ %
\displaystyle{1\over r} <\displaystyle{1\over2}$}.
\end{thm}   

   From this, we easily conclude that 
there are infinitely many tilings in the hyperbolic plane, each one
generated by tessellation from a regular convex polygon~$P$ provided that the number~$p$ 
of sides of~$P$ and the number~$q$ of copies of~$P$ which can be put around a 
point~$A$ and exactly covering a neighbourhood of~$A$ without overlapping satisfy 
the relation:
\hbox{$\displaystyle{1\over p}+\displaystyle{1\over q}<\displaystyle{1\over2}$}.
The numbers $p$ and~$q$ characterize the tiling which is denoted $\{p,q\}$ and
the condition says that the considered polygons live in the hyperbolic plane. Note
that the three tilings of the Euclidean plane which can be defined up to similarities
can be characterized by the relation obtained by replacing~$<$ with~$=$ in the above
expression. We get, in this way, $\{4,4\}$ for the square, $\{3,6\}$ for the equilateral
triangle and $\{6,3\}$ for the regular hexagon.

\vskip 10pt
\setbox110=\hbox{\includegraphics[width=140pt]{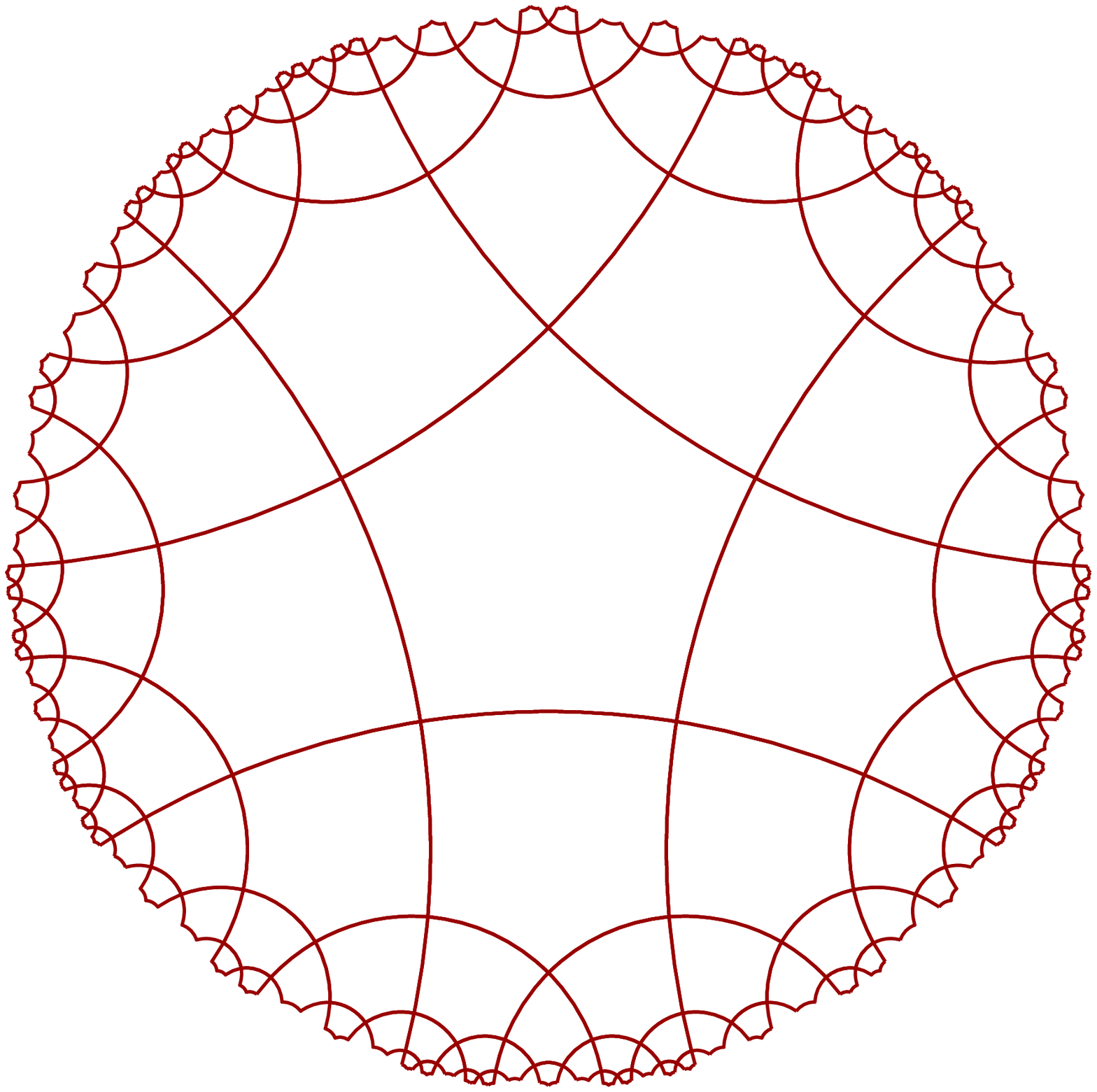}}
\setbox112=\hbox{\includegraphics[width=140pt]{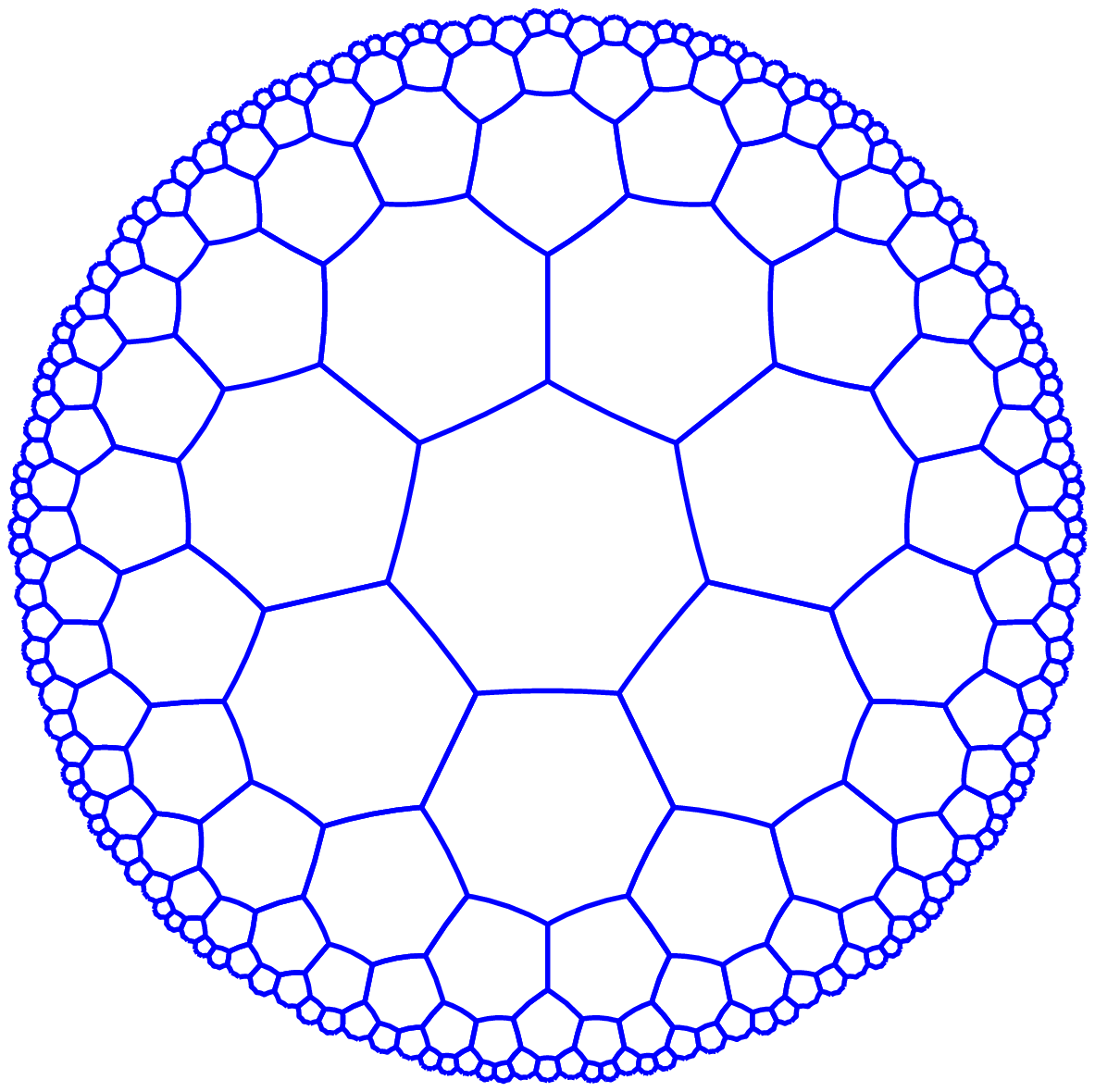}}
\vtop{
\ligne{\hfill
\PlacerEn {-335pt} {0pt} \box110
\PlacerEn {-165pt} {0pt} \box112
}
\vspace{5pt}
\begin{fig} \label{tilings}
\leurre
Left-hand side: the pentagrid. Right-hand side: the heptagrid.
\end{fig}
}

   In the paper, we shall focus our attention on the simplest tilings which can be
defined in this way in the hyperbolic plane: $\{5,4\}$ and~$\{7,3\}$. We call them
the {\bf pentagrid} and the {\bf heptagrid} respectively, see Figure~\ref{tilings}.

   From Figure~\ref{tilings}, it is not clear how to navigate in these tilings, even in
the pentagrid or the heptagrid which are the simplest of the infinite family of
tilings $\{p,q\}$. We introduce such a navigation tool in our next subsection.
As a consequence of the existence of this tool, we can count the tiles of any
tiling $\{p,q\}$.

\subsection{Pentagrid and heptagrid}
\label{GPS}

   The left-hand side picture of Figure~\ref{splitting54} indicates a recursive
splitting of a quarter~$\cal Q$ of the hyperbolic plane which generates the pentagrid.
The idea is that we place a vertex~$V$ of a rectangular regular pentagon~$P_0$ at 
the corner of~$\cal Q$, in such a way that the edges of~$P_0$ which meet at~$V$
are along the sides of~$Q$. These edges are marked {\bf 1} and~{\bf 5} on the figure.
\vskip 10pt
\setbox110=\hbox{\includegraphics[width=140pt]{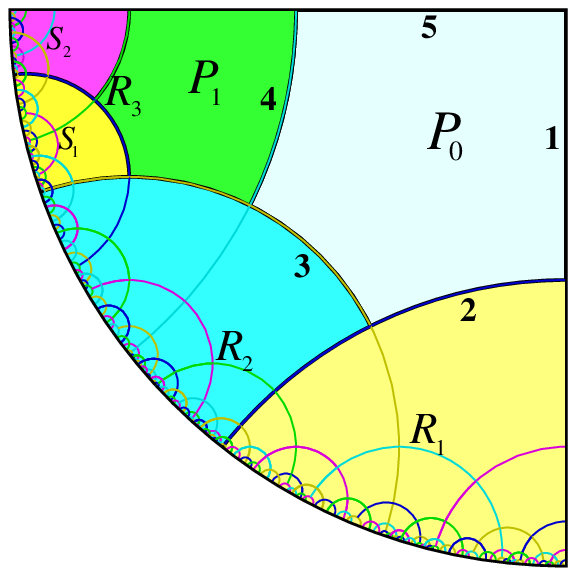}}
\setbox112=\hbox{\includegraphics[width=140pt]{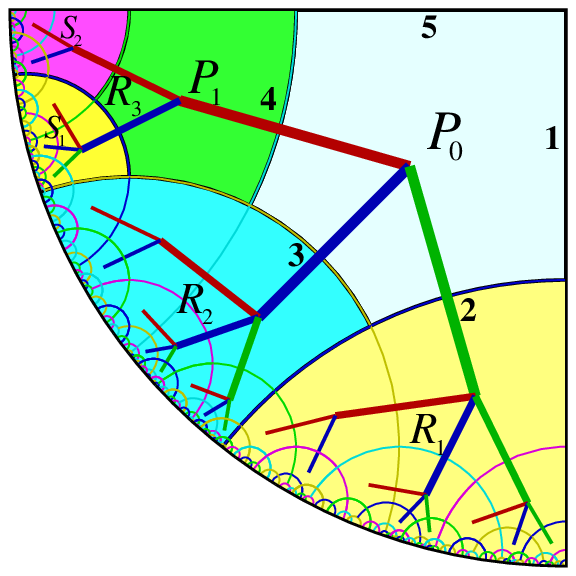}}
\vtop{
\ligne{\hfill
\PlacerEn {-335pt} {0pt} \box110
\PlacerEn {-165pt} {0pt} \box112
}
\vspace{5pt}
\begin{fig} \label{splitting54}
\leurre
The splitting of a quarter of the hyperbolic plane which generates the pentagrid.
On the right-hand side: the tree structure which spans the tiling restricted to the
quarter.
\end{fig}
}

   Next, we consider the complement of~$P_0$ in~$\cal Q$. It can be split in the
three regions labelled $R_1$, $R_2$ and $R_3$ on the figure. The regions $R_1$ 
and~$R_2$ are copies of a quarter: $R_1$ is obtained from~$\cal Q$ by the shift along
side~{\bf 1} moving~$V$ onto the corner of~$R_1$. Similarly, $R_2$ is obtained 
from~$\cal Q$ by the shift along {\bf 4} moving~$V$ on the corner of~$R_2$. What now
remains is $R_3$. Inside~$R_3$, $P_1$ is the reflection of~$P_0$ in~{\bf  4}. The
complement of~$P_1$ in~$R_3$ is split into~$S_1$ and~$S_2$. The region~$S_1$ is again
a quarter obtained from the quarter~${{\cal Q}_1}$ defined by {\bf 5} and~{\bf 4} which 
contains~$P_1$ by the shift along the side of~$P_1$ which is opposite to~{\bf 4}, the 
shift moving the corner of~${\cal Q}_1$ onto that of~$S_1$. Now, it is not difficult
to see that $S_2$ is the image of~$R_3$ by the shift along~{\bf 5} which moves~$V$
onto the corner of~$R_3$. We say that the regions~$R_3$ and~$S_1$ are {\bf strips}.
Now we have two kinds of regions: quarters and strips. Next, Figure~\ref{splitting54}
allows us to split it into quarters and strips again, producing two or three tiles
each time the process is applied to a quarter or a strip. The tree structure
associated to this recursive process is underlined by the right-hand side picture
of Figure~\ref{splitting54}. It is not difficult to see that the tree is
generated by the following rules: $B\rightarrow BW$ and $W\rightarrow BWW$,
considering that the tree has two kind of nodes, {\bf white} and {\bf black} ones.
For each kind of node, the rules indicate what is the kind of its sons and in which
order.

   Now, the tree allows us to number the tiles in a quarter: the root, which
is level~0, receives number~1. Then, on level~1, the tiles receive numbers~2 to~4,
running from left to right on the level. This is repeated on each level. It can be proved
that on the level~$n$, the leftmost tile receives the number $f_{2n}$ and the tiles
on this level are numbered from that number up to $f_{2n+2}$$-1$ which is given to the
rightmost tile, where $f_n$ is the Fibonacci sequence with initial conditions 
\hbox{$f_0=f_1=1$}. This comes from the above rules and the details of the proof 
can be seen in~\cite{mmbook1} where
an important property of this numbering allows us to construct efficient navigation 
tools. Again, we refer to~\cite{mmJUCSii,mmbook1} for the exact proofs and a detailed
account on the navigation. We call {\bf Fibonacci tree} the tree obtained by the
splitting process above described and illustrated by the right-hand side of
Figure~\ref{splitting54}.

\subsection{Extension of the splitting to other tilings}

   This {\it splitting method}, see~\cite{mmDMTCS,mmbook1}, can be extended to
all tilings $\{p,q\}$. We indicate how it works for the heptagrid by 
Figure~\ref{splitting73}, the reader being referred to~\cite{mmDMTCS,mmbook1} for
explanations and more information.

   Let us shortly indicate the basic patterns used in the case of the heptagrid.
This time, we consider an angular sector~$S_0$ defined by the intersection of two rays.
These rays follow special lines adapted to the heptagrid which we call the
{\bf mid-point} lines, as they pass through mid-points of contiguous edges of the tiles.
We consider that the sector contains the tiles which may have at most one vertex outside
the rays with respect to the sector. In Figure~\ref{splitting73}, left-hand side,
the 'big' copy of~$S_0$
defined by the rays $\ell_1$ and~$\ell_2$ and headed by~$\tau$ contains two 'small' 
copies of~$S_0$: the first one is headed by~$\tau_1$ and defined by the rays~$\ell_2$
and~$m_1$; the second one is headed by~$\tau_2$ and defined by the rays~$m_1$
and~$m_2$. We have a second region, $S_1$, which we again call a {\bf strip}: it is
headed by~$\tau_3$ and defined by the rays~$m_2$ and~$\ell_1$. Note that the lines
which support these rays are non secant. It is not difficult to see that the line
supporting the edge $s_3$ is their common perpendicular. Note that the copy of~$S_0$
headed by~$\tau_1$ is obtained from the copy of~$S_0$ headed by~$\tau$ by the 
shift~$\sigma$ illustrated by the read arrow of the left-hand side of 
Figure~\ref{splitting73}.

   Now, it is not difficult to see that $S_0$ is spanned by the same tree as the tree
spanning~$\cal Q$, see the right-hand side of Figure~\ref{splitting73}.

   This identical spanning tree for the pentagrid and for the heptagrid is not
a particular feature of both these tilings. It can be generalized to an infinite
family of tilings, the tilings $\{p,4\}$ and $\{p$+$2,3\}$, with $p\geq 5$. The tilings
$\{p,4\}$ consists of the tessellations based on a regular rectangular polygon
while the tilings $\{p$+$2,3\}$ consists of those based on a regular polygon with
angle $\displaystyle{{4\hbox{\bf d}}\over 3}$, {\bf d} being the measure of the
right angle. We have that for each $p$
with $p\geq 5$, the same tree spans the tilings $\{p,4\}$ and $\{p$+$2,3\}$. Of course,
for $p>5$ the tree is no more the Fibonacci tree: it is another tree, connected with
another recurrent sequence, associated with an algebraic number, 
see~\cite{mmDMTCS,mmbook1}. 
 
   We would like to remark that, as proved in~\cite{mmbook1}, the splitting method
can also be applied to all tilings $\{p,q\}$ and provides a tree which spans the
tiling. This tree is more complex than the ones we devised for the tilings $\{p,4\}$
or $\{p$+$2,3\}$.

\setbox110=\hbox{\includegraphics[width=170pt]{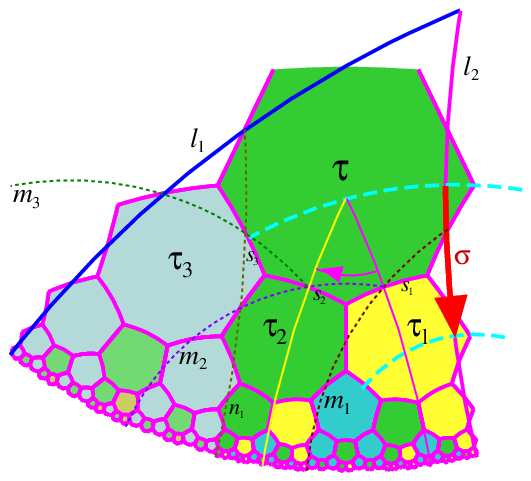}}
\setbox112=\hbox{\includegraphics[width=170pt]{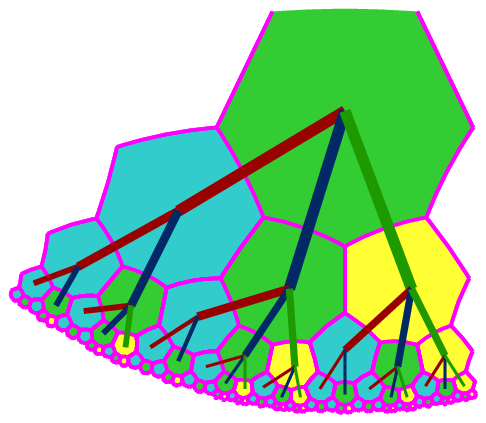}}
\vtop{
\ligne{\hfill
\PlacerEn {-330pt} {0pt} \box110
\PlacerEn {-175pt} {0pt} \box112
}
\vspace{0pt}
\begin{fig} \label{splitting73}
\leurre
The splitting of a sector of the hyperbolic plane which generates the heptagrid. 
Notice the underlying tree, the same one as for the pentagrid.
\end{fig}
}

   Now, we have all elements in order to see how the infinite numeral system can
bring in a more precise information to the picture which was described in this section.
Before turning to this study, let us mention a few applications of the pentagrid and
the heptagrid. Among those quoted in~\cite{mmbook2}, let us mention the
colour chooser using the heptagrid, see~\cite{palette}, the Japanese keyboard for
cellphones using the pentagrid, see~\cite{acri2} and the communication protocol
between tiles of the pentagrid or the heptagrid, see~\cite{mmJCAcomm,mmHongKong}.

\section{{\Large Even and odd splittings and their applications}}
\label{hypgrossone}

   First, we look at the application of the infinite numeral system to the pentagrid
and to the heptagrid. In Subsection~\ref{generalization}, we shall see how to
generalize these results to the family of tilings $\{p,4\}$ and $\{p$+$2,3\}$, and in
Section~\ref{generalcase}, we look at the application of the infinite numeral system
to the tilings $\{p,q\}$.

\subsection{Pentagrid and heptagrid}
\label{pentagrossone}

   In the case of the pentagrid, it is easy to see that we have two ways to realize the
tiling. One way is based on the observation that the whole tiling can be split into 
exactly four quarters as illustrated on the left-hand side picture of 
Figure~\ref{evensplittings}. Call this way the {\bf even splitting}. 

The other way consists in choosing a tile which will be called the {\bf central tile} 
and then to notice that the complement of the tile in the plane can be split into 
exactly five quarters, see the left-hand side of Figure~\ref{oddsplittings}. Call this
second way the {\bf odd splitting}. Both ways are thoroughly explained in~\cite{mmbook1}. 

   Now, we can see that each quarter has exactly the same number of tiles as
they are copies of each other which is based on a geometrical property.
But what is this number of tiles? In~\cite{mmIW}, it was said that,
'from the enumeration property mentioned in Subsection~\ref{GPS}, we can see that
the number of tiles contained in each quarter is \grossone{}'. We can consider this
estimation as a {\it first approximation} of the expected number of tiles.
This first approximation shows the same characteristics as the rule on
infinity in traditional calculus: $\infty+\infty=\infty$. As remarked
by Yaroslav Sergeyev in~\cite{YSTuring}, this is analogous to the rule
\hbox{'many' + 'many' = 'many'} in Pirah$\tilde{\hbox{\rm a}}$'s numeral
system consisting of 1, 2 and~'many', see the above paper and~\cite{gordon}
for more information on Pirah$\tilde{\hbox{\rm a}}$, a primitive tribe living
in Amazonia.

Let us look closer at what happens and how the infinite numeral system
can give us a more precise information. 

In Subsection~\ref{GPS}, we indicated
that the number of nodes of the Fibonacci tree which stand at the level~$n$
is $f_{2n+1}$. From this, it is not difficult to prove that
the number of nodes of the Fibonacci tree which are on a level~$m$ with $m\leq n$
is~$f_{2n+2}$$-$$1$. Now, the Fibonacci tree is an infinite tree and, as it has
a finite bounded branching, its height is also infinite. As we can assign a number
to each tile of a quarter of the pentagrid, it is reasonable to assume that
we have at most \grossone{} tiles in a quarter. This leads us to consider
the set~$\cal F$ of numbers~$n$ such that $f_n\leq\grossone$, extending the
Fibonacci sequence to infinite indices by simply assuming that the
induction definition $f_{n+2} = f_{n+1}+f_n$ still applies to infinite indices
and to the consequently infinite terms of the extended sequence. Notice
that explicit values can be obtained by using the expression:
$f_x = \displaystyle{1\over{\sqrt5}}\Big(\big(\displaystyle{{1+\sqrt5}\over2}\big)^x - 
\big(\displaystyle{{1-\sqrt5}\over2}\big)^x\Big)$,
where $x$~can take any positive integer values, infinite ones being included.
Define~$\mu$ as the number of elements of~$\cal F$. We know that we can 
consider~$\mu$ as the greatest element of~$\cal F$. This gives us that 
$f_\mu\leq \grossone < f_{\mu+1}$. Consequently, if 
$\vartheta=\displaystyle{{1+\sqrt5}\over2}$, we have that
$\mu\asymp \log_\vartheta\grossone$. 

   Now, let us remark that if we apply the rules defining the Fibonacci tree
to a black node, we obtain another tree, which we call the {\bf smaller} Fibonacci
tree. By contrast, we call the Fibonacci tree issued from a white node
the {\bf standard} Fibonacci tree. It is not difficult, using the rules which 
define a Fibonacci tree, to prove that the number of nodes on the level~$n$ of a 
smaller Fibonacci 
is $f_{2n}$, so that the number of nodes of the level~$m$ with $m\leq n$ 
is~$f_{2n+1}$.

   Now, as the standard Fibonacci tree is clearly a faithful realization of the 
properties of the Fibonacci sequence, we can see that a smaller Fibonacci tree is 
indeed included in a standard one of the same height: the inclusion is even true
at each level. We may assume that once the counting of a 
level~$x$ is performed, it goes to the end of level~$x$: if the first nodes of 
level~$x$ have sons, there is no reason to consider that the last nodes of the same
level have no son. These remarks have the following impact on~$\mu$. 
As a consequence, if $\mu$ were odd, a smaller Fibonacci tree
of height~$\mu$ could be realized by not a standard Fibonacci tree of the same height,
a contradiction with the previous assumption. Accordingly, we may assume that
$\mu$~is even, so that we can write $\mu=2\eta$.  

\vskip 5pt
\setbox110=\hbox{\includegraphics[width=140pt]{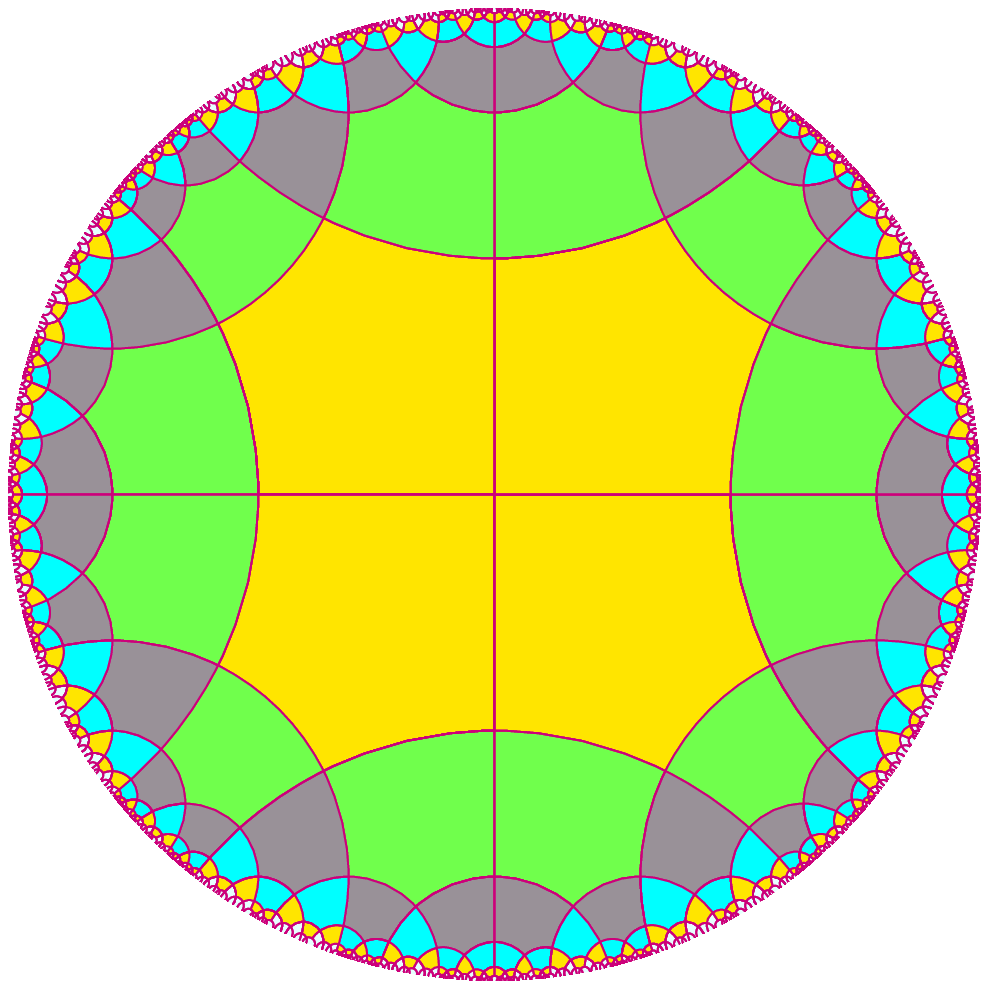}}
\setbox112=\hbox{\includegraphics[width=140pt]{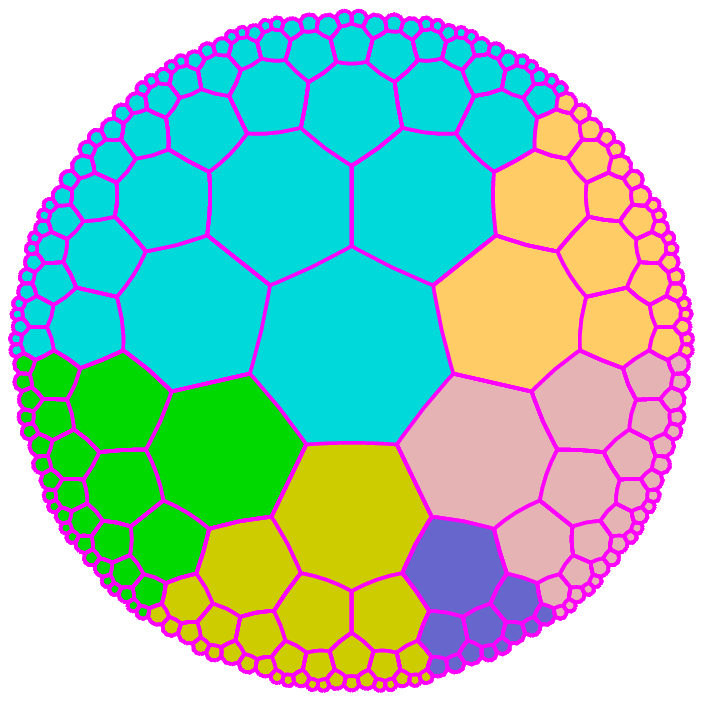}}
\vtop{
\ligne{\hfill
\PlacerEn {-330pt} {0pt} \box110
\PlacerEn {-165pt} {0pt} \box112
}
\vspace{-5pt}
\begin{fig} \label{evensplittings}
\leurre
The even splitting in the pentagrid, left-hand side, and in the heptagrid, 
right-hand side.
\end{fig}
}

Accordingly, we can consider that the number of tiles in a quarter of the pentagrid
is $W = f_{\mu}$$-$1, so that the eight of the spanning tree of the quarter 
is~$\eta$. And so, in the even splitting of the pentagrid
we get $4W$~tiles, while we get 1$+5W$~tiles in the odd splitting.
This is indicated in Theorem~\ref{numberpentahepta} and in
Table~\ref{table1}.

\vskip 5pt
\setbox110=\hbox{\includegraphics[width=140pt]{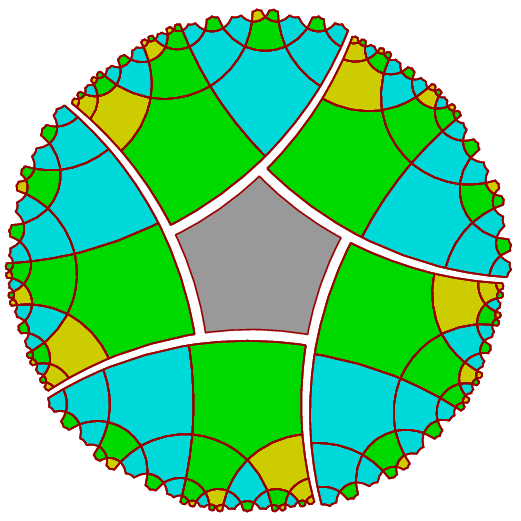}}
\setbox112=\hbox{\includegraphics[width=140pt]{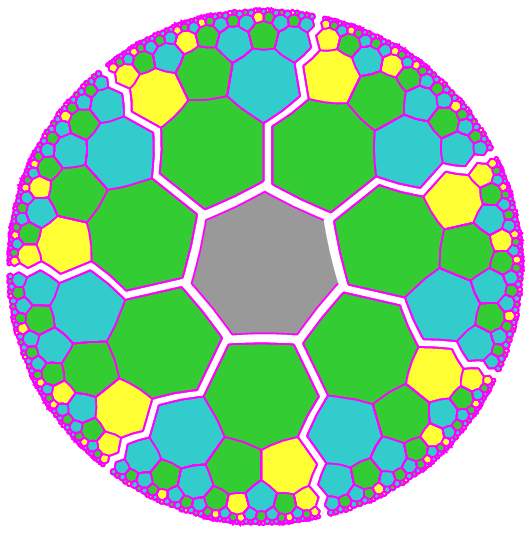}}
\vtop{
\ligne{\hfill
\PlacerEn {-330pt} {0pt} \box110
\PlacerEn {-165pt} {0pt} \box112
}
\vspace{-5pt}
\begin{fig} \label{oddsplittings}
\leurre
The odd splitting in the pentagrid, left-hand side, and in the heptagrid, right-hand
side.
\end{fig}
}

   In the case of the heptagrid, the odd splitting can easily be established. It
is illustrated by the right-hand side picture of Figure~\ref{oddsplittings}.
This gives us 1$+7W$~tiles.

   The even splitting can also be defined for the heptagrid. We display three sectors
whose heading tiles share a common vertex~$V$, as illustrated by the right-hand side
picture of Figure~\ref{evensplittings}. Now, we have to take into account the
heights of the Fibonacci trees involved in the splitting, which are not the same
for each part. In order to make things comparable, we cannot take~$V$ as a centre,
as we have no simple geometrical transformation which would allow us to compare the
sectors. And so, instead of~$V$, we take as the centre, one of the tiles which has~$V$
among its vertices, say~$T$. We may consider that $T$~is the light blue cell in 
the right-hand side picture of Figure~\ref{evensplittings}, also the central cell of 
the figure. The sector headed by~$T$ 
contains 1+$2W$+$B$ tiles, where $B$~is the number of tiles spanned by a smaller 
Fibonacci tree. We can easily see, that $W$~is the number of tiles of the four other
sectors headed by a tile which shares an edge with~$T$. We remain with a sector
whose height is~$\eta$$-$1. Let $W_1$ be the number of tiles spanned by a 
standard Fibonacci tree of height~$\eta$$-$1. Note that $W_1 = f_{2\eta-1}$$-$1
and that $B=f_{2\eta}$, so that $W_1+B=f_{2\eta+1}-1=W$. Accordingly, we find again
1$+7W$~tiles.

\vtop{
\begin{tab} \label{table1}
\leurre
Table of the total area observable through the even and the odd splittings for the
pentagrid and for the heptagrid. Remember that {\bf d} is the measure of the right angle.
\end{tab}
\vspace{-14pt}
\grostrait
\ligne{\ttV\hskip 40pt
\hbox to 80pt{\hfill}
\hbox to 80pt{\hfill pentagrid\hfill}
\hbox to 80pt{\hfill heptagrid\hfill}\hfill}
\vspace{2pt}
\demitrait
\vspace{2pt}
\ligne{\ttV\hskip 40pt
\hbox to 80pt{even splitting\hfill}
\hbox to 80pt{\hfill $4W\cdot\hbox{\bf d}$\hfill}
\hbox to 80pt{\hfill $\displaystyle{{14}\over3}W\cdot\hbox{\bf d}$%
+$\displaystyle{2\over3}${\bf d}\hfill}\hfill}
\ligne{\ttV\hskip 40pt
\hbox to 80pt{odd splitting\hfill}
\hbox to 80pt{\hfill $5W\cdot\hbox{\bf d}$+{\bf d}\hfill}
\hbox to 80pt{\hfill $\displaystyle{{14}\over3}W\cdot\hbox{\bf d}
$+$\displaystyle{2\over3}${\bf d}\hfill}\hfill}
\demitrait
\vspace{10pt}
}
   And so, we find a difference with the estimation of~\cite{mmIW} which was based
on a less precise estimation on the number of tiles. As a consequence, the number of 
tiles do not lead to the same observable area in the case of the pentagrid and in
the case of the heptagrid.  

\begin{thm}\label{numberpentahepta}
Let $W$~be the number of tiles in a standard Fibonacci tree of height~$\eta$
where $f_{2\eta}\leq\grossone < f_{2\eta+1}$.
The number of tiles of the pentagrid which can be observed with the
help 
of the infinite numeral system 
is of the form $4W$
when using the even splitting while it is of the form 
$1$$+$$5W$~in the odd splitting. In the case of the heptagrid, the number of tiles
is the same under the even or the odd splitting, and it is $1$$+$$7W$ in both cases.
The total area covered by the tiles is given by the following table, see
Table{\rm~\ref{table1}}. Denote by $P_e$, $P_o$ the total area which is observable 
in the pentagrid under the even, odd splitting respectively. Denote by $H$ the
total area which is observable in the heptagrid. Then we have 
that \hbox{$P_e< H < P_o$} and, more precisely, \hbox{$H-P_e = 2(P_o-H)$}.
\end{thm} 

  Now, we turn to the total area which can be observed under each splitting.

\noindent
Proof.
Remember that the area of a triangle is the complement to the sum of the angles in order
to obtain a straight angle. Remember that a straight angle is two right angles. We denote
the right angle by {\bf d}. From this, we obtain that the area of a regular rectangle
pentagon is \hbox{3.2{\bf d} $-$ 5.{\bf d}} and so it is {\bf d}. Similarly,
the area of a regular heptagon with angle $\displaystyle{{4\hbox{\bf d}}\over 3}$ 
is \hbox{5.2{\bf d} $-$ 7.$\displaystyle{{4\hbox{\bf d}}\over3}$}, so that it is
$\displaystyle{{2\hbox{\bf d}}\over3}$. From this we obtain the values indicated
in Table~\ref{table1}. The values of~$P_e$, $P_o$ and~$H$ easily lead to
the relation \hbox{$H-P_e = 2(P_o-H)$}, as the usual operations on numbers also
hold for the infinite ones. This completes the proof of the theorem. \cqfd


   It is interesting to notice that the total area which is observable in the
case of the heptagrid is in between the total area in the case of the pentagrid
under the even or the odd splitting. It is also interesting to notice that
the odd splitting provides a significantly larger area in the case of the pentagrid,
although the number of tiles is significantly less.
This corresponds to the fact that the area of the regular rectangle pentagon is 
significantly bigger than that of the heptagon with the angle 
$\displaystyle{{4\hbox{\bf d}}\over3}$.

  At last, it is worth noticing that the numbers of tiles $4W$ and $5W$+1 and $7W$+1
which we obtained are all bigger than~$\grossone$. This easily comes from the
fact that $f_n<3f_{n+1}$ which is generalized to infinite indices and from
the definition of $W=U_{2\eta}$. From the very definition of $\mu=2\eta$,
we have that $U_\mu\leq\grossone < U_{\mu+1}$ and, clearly, $U_{\mu+1}<3U_\mu=3W$.
This property will turn out to be true in all the situations we consider in the paper.

   Before turning to the generalization of the result of Theorem~\ref{numberpentahepta}
to other tilings of the same family, we first remark that we obtain the same results
if we change the place of the central tile in an odd splitting or of the central vertex
in an even splitting. The second remark is an interesting generalization
of the odd splitting which gives similar results\footnote{It is my pleasure to thank 
Yaroslav Sergeyev for two questions he raised after reading the initial version of 
the paper. It is now the occasion to give the answers.}.

\vskip 5pt
\setbox110=\hbox{\includegraphics[width=180pt]{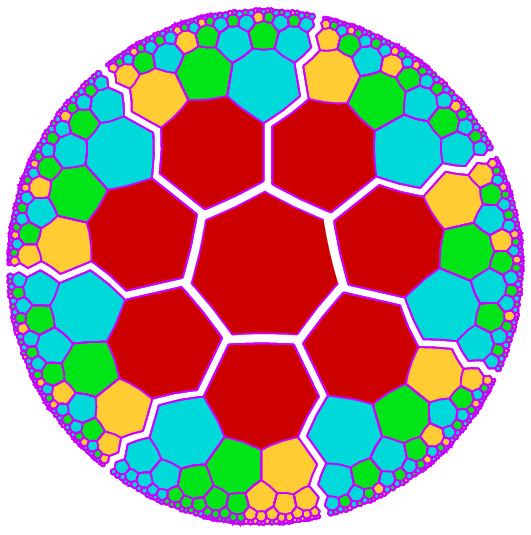}}
\vtop{
\ligne{\hfill
\PlacerEn {-280pt} {0pt} \box110
}
\vspace{0pt}
\begin{fig} \label{oddsplittingball}
\leurre
Illustration, in the heptagrid, of the principle of the considered generalization:
fixing a ball around the central tile, the tiles in red, the darkest tiles of the
figure. Here, the radius of the ball is~$1$.
\end{fig}
}
\vskip 10pt
   Figure~\ref{oddsplittingball} shows the principle of a family of odd splittings
which generalizes the one which we performed. The figure illustrates a particular
case in the heptagrid. The general definition, which applies to both the pentagrid
and the pentagrid is as follows. 

First, we say that a tile~$A$ is at a {\bf distance}~$k$ in tiles from a tile~$B$,
if there is a sequence \hbox{$T_0$, ..., $T_h$} of tiles with $T_0=A$, $T_h=B$,
$T_i$ and~$T_{i+1}$ share a side for $i\in\{0..k$$-$$1\}$ such that $k=h$
and that there is no such sequence for $h < k$. 
Then, from this we define a {\bf ball of radius~$r$ around a tile~$T$}, with~$r$ 
a positive number, as the set of tiles whose distance in tiles from~$T$ is at most~$r$. 

Now, we fix a ball of radius~$r$ around the central tile. 
And then, we split what remains in the space into sectors of the two kinds defined 
in Figures~\ref{splitting54} and~\ref{splitting73}. The possibility to perform 
such a splitting is proved in~\cite{mmTUCS,mmbook1}. From these studies, we know that
the number of tiles in the ball is 5$f_{2r-1}$+1{} in the pentagrid and 7$f_{2r-1}$+1{} 
in the heptagrid. Now, the number of sectors we need, is defined by the set of tiles
which are outside the ball and also in contact with a tile of the ball. The number
of these latter tiles is 5$f_{2r+1}$ in the pentagrid and 7$f_{2r+1}$. We have two
kind of sectors which correspond to the one defined Figures~\ref{splitting54} 
and~\ref{splitting73}, say white, black ones spanned by a standard, smaller Fibonacci 
tree respectively. As each node produces one black son exactly, the number of
black nodes on the level~$n$+1 of a standard, smaller Fibonacci tree is the number
of nodes on the level~$n$ and so it is $f_{2n+1}$, $f_{2n-2}$ respectively. This gives
us $f_{2n+2}$, $f_{2n-1}$ white nodes respectively.

We know the number of tiles spanned by each kind of sector. However, we do not 
start from a sector whose heading tile shares a side with the central cell but 
at a distance~$r$ from it. Accordingly, the height of such a sector is no more~$\eta$, 
it is $\eta$$-$$r$. Accordingly, each white sector contains $W_r=f_{2(\eta-r)+1}-1$ 
tiles and each black one contains $B_r=f_{2(\eta-r)}$ of them. 
Note that this is conformal to Postulate~3, as clearly a sector spanned by a
standard Fibonacci tree strictly contains isometric images of itself. The
definition of the height of such a sector allows us to measure precisely the
difference with the central sector. 

Now, call such an odd splitting an 
{\bf $r$-augmented odd splitting}. This allows us to state the following 
result:

\begin{thm}\label{r_augmented}
Under an $r$-augmented odd splitting, the number of tiles of the pentagrid which 
can be observed with 
the infinite numeral system 
is given by: 
$5f_{2r-1}B_r$$+$$5f_{2r}W_r$$+$$5(f_{2r}-1)$$+$$1$ for the pentagrid and
for the heptagrid, it is given by:
$7f_{2r-1}B_r$$+$$7f_{2r}W_r$$+$$7(f_{2r}-1)$$+$$1$.
Accordingly, the observable total
area is 
$(5f_{2r-1}B_r$$+$$5f_{2r}W_r)${\bf d}%
$+$$(5(f_{2r}-1)$$+$$1)${\bf d} for the pentagrid and
in the heptagrid it is given by: 
$\displaystyle{{14}\over3}(f_{2r-1}B_r$$+$%
$f_{2r}W_r)${\bf d}%
$+$$\displaystyle{{14}\over3}(f_{2r}-1)${\bf d}$+$$\displaystyle{2\over3}${\bf d}. 
\end{thm}

   Note that when $r=0$, and setting $f_1=0$, we find again the results of 
Theorem~\ref{numberpentahepta}.

   Theorem~\ref{r_augmented} shows that, in this case too, the total area which is 
observable in the pentagrid is greater than that which is observable in the heptagrid 
although more tiles are observable in the latter than in the former. 

\subsection{General scheme for the splittings $\{p,q\}$}
\label{generalscheme}

   In the next sub-sections, we shall always follow the same scheme
which can be formulated as follows.

   First, we remind the splitting of the tiling. At this stage, we shall see
that the general case can be split into several general situations:
the tilings $\{p,4\}$ and $\{p$$+$$2,3)\}$ which generalize most properties
of the pentagrid and the heptagrid, including the connection between this tilings;
the tilings $\{p,q\}$ when $q$~is even and then, when $q$~is odd. We shall see that
for this latter case, we have two solutions.
 
   In this splitting, a region~$S_0$ will play the most important role: we shall call
it the {\bf basic region}, and its spanning tree, the tree of the splitting will
be denoted by~$\cal T$. In all cases, the splitting induces a polynomial of the 
splitting from which we deduce a recurrent polynomial equation satisfied by the 
number~$_u$ of nodes which are on the level~$n$ of~$\cal T$. Of course, we shall
extend the recurrent equation to infinite indices and, consequently, to the infinite
values of $u_n$ corresponding to infinite~$n$'s. We shall then consider
the number $U_n$~of nodes on the levels~$m$ of~$\cal T$ with $m\leq n$.
We know that $U_n=\displaystyle{\sum\limits_{i=0}^nu_n}$.
With the help of these numbers, also extended to infinite integral indices,
we can define the number~$\eta$ of positive integers~$x$, finite of infinite,
such that $U_x\leq\grossone$. This number will be considered as the height of~$\cal T$. 
We shall immediately get that the number of tiles observable in~$S_0$
is given by~$W=U_\eta$. There will be no inconvenience to denote by the same letter~$W$
the different infinite numbers denoted bu $U_\eta$, where $U_x$ refers to the 
sequence~$u_n$, $u_n$ following different equations, depending on the splitting
considered for the same tilings or for different ones. In each sub-section or
sub-subsection, the meaning of~$S_0$, $\cal T$, $u_n$, $U_n$, $\eta$ and~$W$
will be the same. Thanks to~$W$ and to the splitting of the tiling, it will be possible
for us to give an expression for the number of tiles which are observable
with the infinite numeral system based on~\grossone.

   Also, we shall compute the area $\alpha$~of the regular polygon on which the
tiling is constructed, which we shall call the {\bf basic polygon}. From this
computation and from the area of a tile, we shall be able to compute a 
precise expression for the observable total area. This will allow us to compare
this area for various splittings of the same tilings or for different tilings when this
comparison will make sense.

\subsection{Generalization to $\{p,4\}$ and to $\{p$$+$$2,3\}$}
\label{generalization}

   This sub-section is more different than what was presented in~\cite{mmIW}.

   As announced in the Sub-section~\ref{generalscheme}, this sub-section gives
a direct generalization of what was done in the case of the pentagrid and of
the heptagrid. The important common point with Subsection~\ref{pentagrossone} is
that the splittings of the tilings~$\{p,4\}$ and~$\{p$$+$$2,3\}$, with $p\geq5$, 
are spanned by the same tree. It is no more the Fibonacci tree, but the new tree 
has several common properties with the Fibonacci tree.

   We have again two kinds of nodes, corresponding to the two kinds of regions involved
in the splitting: again we call them {\bf black} and {\bf white}, denoted by
$B$ and~$W$ respectively, the white nodes
being attached to the bigger region. The tree which spans the big region is called
{\bf standard} and that which is spans the small region is called {\bf smaller}
and the standard tree strictly contains the smaller tree. The rules are now
\hbox{$W\rightarrow BW^{p-3}$} and \hbox{$W\rightarrow BW^{p-4}$}, with 
\hbox{$p\geq5$}.
This allows to easily infer the splitting of the tiling in these cases.

As proved in~\cite{mmbook1}, the polynomial of the splitting
is \hbox{$X^2- (p$$-$$2)X + 1$} and so, the number~$u_n$ of nodes on the 
level~$n$ of the tree satisfies the following recurrent equation:
\hbox{$u_{n+2} = (p$$-$$2)u_{n+1} - u_n$}. The sequence is increasing with~$n$
and we have that $u_n \asymp \beta^n$, with 
$\beta =\displaystyle{{p-2+\sqrt{p(p-4)}}\over2}$.
For a standard tree, we have \hbox{$u_{-1}=0$} and \hbox{$u_0=1$}, and for a smaller 
tree, we have \hbox{$u_{-1}=u_0=1$}.
We extend the sequence to any integer, finite or infinite by the recurrent equation.
Note that from the recurrent equation, using the standard representation involving
a square $2\times2$-matrix, and $\beta$~being the biggest real root of
\hbox{$X^2-(p$$-$$2)X+1$}, we can prove the following expression of~$u_n$:
$u_n=\displaystyle{\beta\over{\beta^2-1}}\Big(\beta^{n+1}-%
\displaystyle{1\over{\beta^{n+1}}}\Big)$, where $n$~is a natural number, the
result of the computation being also a positive integer. Extending $n$~to positive 
infinite integers, this formula allows us to give an expression for~$u_n$
when it is infinite.

Together with~$W$, we have to estimate the number of nodes of another tree~$\cal A$: 
the one which spans the other region of the splitting which is called~${\cal S}_1$,
see~\cite{mmbook1}. This time, we denote by~$v_n$ the number of nodes in the 
level~$n$ of~$\cal A$ and by~$V_n$, the number of nodes in the levels~$m$ with 
$m\leq n$ of~$\cal A$. We know that $v_n$~follows the same recurrent equation 
as~$u_n$ with different initial values: $v_0=1$ and $v_1=u_1$$-$1. We also
know that $V_m < V_n$, and we denote by~$B$ the number $V_{\varphi_p}$,
observing that $B < W$ as, clearly, ${\cal A}\subset \cal T$, the inclusion being 
proper.

Now, we can see that
for the tiling $\{p,4\}$, we have a sharp difference between the even splitting
which provides us with $4W$~tiles and the odd one which yields $1$+$pW$~tiles.
Now, in the case of the tiling $\{p$+$2,3\}$, this time we again shall find
the same number of tiles for both the even and the odd splittings.

Indeed, the decomposition we observed for the even splitting of the 
heptagrid works word by word. For the same reason, we fix the central cell~$C$ as
one of the three tiles heading a sector which share a common vertex. Then, around~$C$, 
we have $p$$-$1~sectors spanned by the standard tree, one spanned by the standard tree
of height~$\varphi_p$$-$1, which gives $W_1=U_{\varphi_p-1}$ tiles. Now, the sector
which is head by~$C$ also contains a region which is spanned by the smaller tree, for
the same reason as in the heptagrid. This gives $B$~tiles which, added to the 
$W_1$~tiles already mentioned give us $W$~tiles as $W = B+W_1$, which is geometrically
clear from Figure~\ref{evensplittings}: indeed, the smaller region can be obtained
from the smaller region by deleting a copy of the standard one. This can easily be seen
on the spanning tree: the root of the standard tree has $p$$-$2 sons while the
root of the smaller tree has $p$$-$3 of them, and the difference is a white tree.
Accordingly, we obtain 1+$(p$+$2)W$~tiles in the even splitting of the 
tiling $\{p$+$2,3\}$. 
In the odd splitting of the same tiling, we again find 1+$(p$+$2)W$ tiles for the same
reason as in the heptagrid. And so, we can see that the situation which we observed in
the heptagrid is generalized to all the tilings $\{p$+$2,3\}$.

   Now, we have to compute the observable total area in all cases.

In the general case of a regular polygon with $p$~sides and of vertex angle
$\displaystyle{{2\pi}\over q}$, splitting the area of the polygon into 
$p$$-$2~triangles, we find that the area is \hbox{$(p$$-$2$)2${\bf d}}
minus the sum of the interior angles, {\it i.e.} 
$p.\displaystyle{{2\pi}\over q}=p.\displaystyle{{4\hbox{\bf d}}\over q}$.
When $q=4$, this gives us \hbox{$(p$$-$$4)${\bf d}}. When $q=3$, with a regular polygon
with $p$+2 sides, this gives us \hbox{$(p$$-$$4)\displaystyle{{2\hbox{\bf d}}\over3}$}.
This shows us that again, the  area of the regular rectangular polygon with $p$~sides 
is bigger than the regular polygon with the angle 
$\displaystyle{{4\hbox{\bf d}}\over3}$ with $p$+2 sides. Despite the bigger
number of tiles in the tiling $\{p$$+$$2,3\}$, the total area is still bigger for
the tiling $\{p,4\}$ under the odd splitting, as we can see from 
Table~\ref{table2}. As in the
case of the heptagrid and the pentagrid, the total area for the tiling
$\{p$+$2,3\}$ is the same for both splittings and it takes an intermediate
value between the total area for the tiling $\{p,4\}$ under the even splitting
and its total area under the odd splitting. This time, the total area of the
tiling $\{p$+$2,3\}$ is much closer to the total area of the tiling $\{p,4\}$
under the odd splitting.

   
\begin{thm}\label{numberp4pp3}
In the tiling~$\{p,4\}$, the number of tiles which can be observed with the infinite
numeral system is given by $4W$~under the even splitting and by $1$$+$$pW$~under the
odd splitting. For the tiling~$\{p$$+$$2,3\}$, the corresponding number of tiles
is $1$$+$$(p$$+$$2)W$ under both the even and the odd splittings. The total area 
which can be observed in each case is given by Table~{\rm\ref{table2}}. Denote
by $P_e$, $P_o$ the total area which is observable in the tiling $\{p,4\}$ under
the even, odd splitting respectively and denote by $H$ the observable total area
in the tiling $\{p$$+$$2,3\}$. Then, we have that
$H-P_e=2(P_o-H)$.
\end{thm} 

Indeed, the computation which we can perform from Table~\ref{table2} gives us that
\hbox{$H-P_e = \displaystyle{2\over3}(p$$-$$4)^2W\hbox{\bf d} 
+ \displaystyle{2\over3}(p$$-$$4)\hbox{\bf d}$}
and \hbox{$P_o-H= \displaystyle{1\over3}(p$$-$$4)^2W\hbox{\bf d}
+ \displaystyle{1\over3}(p$$-$$4)\hbox{\bf d}$}. 

This completes the proof of
Theorem~\ref{numberp4pp3}. \cqfd

\vtop{
\begin{tab} \label{table2}
\leurre
Table of the total area observable through the even and the odd splittings for the
tilings $\{p,4\}$ and $\{p$$+$$2,3\}$ tilings, {\bf d} being the measure of the 
right angle.
\end{tab}
\vspace{-14pt}
\grostrait
\ligne{\ttV\hfill 
\hbox to 60pt{\hfill}
\hbox to 110pt{\hfill $\{p,4\}$\hfill}
\hbox to 130pt{\hfill $\{p$$+$$2,3\}$\hfill}\hfill}
\vspace{2pt}
\demitrait
\vspace{2pt}
\ligne{\ttV\hfill 
\hbox to 60pt{even splitting\hfill}
\hbox to 110pt{\hfill $4(p$$-$$4)W$.{\bf d}\hfill}
\hbox to 130pt{\hfill $(p$$+$$2)(p$$-$$4)W{}\displaystyle{{2\hbox{\bf d}}\over3}$%
$+$$(p$$-$$4)\displaystyle{{2\hbox{\bf d}}\over3}$\hfill}\hfill}
\ligne{\ttV\hfill 
\hbox to 60pt{odd splitting\hfill}
\hbox to 110pt{\hfill $p(p$$-$$4)W$.{\bf d}$+$$(p$$-$$4)${\bf d}\hfill}
\hbox to 130pt{\hfill $(p$$+$$2)(p$$-$$4)W{}\displaystyle{{2\hbox{\bf d}}\over3}$%
$+$$(p$$-$$4)\displaystyle{{2\hbox{\bf d}}\over3}$\hfill}\hfill}
\demitrait
}
\vskip 10pt
  At last, and not the least, it is also possible in this case to define an 
$r$-augmented odd splitting, both in the tilings $\{p,4\}$ and $\{p$$+$$2,3\}$, 
considering a ball of radius~$r$ around the central cell. However, the estimation of 
the number of observable tiles and their areas involve developments of~\cite{mmbook1} 
which we have no room to reproduce here.

\subsection{The general case $\{p,q\}$}
\label{generalcase}

   This case is more complex than the case of the tilings studied in 
Subsection~\ref{generalization} by the fact that the splittings in the
case when $q$~is even and in the case when it is odd are very different.
Moreover, there is no simple way to define $r$-augmented odd splittings.
And so, we shall look at the even and odd splittings only.

First, in Sub subsection~\ref{generaleven}, we remind the splitting of~\cite{mmbook1} 
in the case when $q$~is even. Then, in Sub subsections~\ref{generalodd1}
and~\ref{generalodd2}, we deal with the case when $q$~is odd. As announced in
Sub-section~\ref{generalscheme}, we shall offer two different splittings
when $q$~is odd, each one having its own merit.

\subsubsection{The case when $q$ is even}
\label{generaleven}

In that case, we define a {\bf sector}~${\cal S}_0$ as the angular 
sector defined by taking a vertex~$V$ of the polygon~$P$ on which the tiling is 
constructed and the rays issued from~$V$ which supports the two edges of~$P$ which 
meet at~$V$. We call $V$ the vertex of~${\cal S}_0$ and $P$ is its {\bf head}. It is 
easy to see that the whole tiling is the union of $q$~copies of~${\cal S}_0$ which 
share the same vertex.

\ifnum 1=0 {
Inside ${\cal S}_0$, consider the edge~$e_1$ of~$P$ which is supported by the left-hand 
side ray~$\rho_\ell$ defining ${\cal S}_0$. Define $e_i$ with $i\in\{2..p\}$ as the other 
edges of~$P$, numbering them from~$e_1$ while counter clockwise turning around~$P$. Define
$V_i$, $i\in\{1..p\}$ to be the vertices of~$P$, $V_i$ and~$V_{i+1}$ being the ends 
of~$e_i$ for $1\leq i< p$ and $V_p$ and $V_1$ being the ends of~$e_1$. As $q$~is 
even, we can write $q=2h$ and we can easily see that the angle between $e_2$ 
and~$\rho_r$ which is outside~$P$ and inside ${\cal S}_0$ is 
$(h$$-$$1)\displaystyle{{4\hbox{\bf d}}\over q}$, see~Figure~\ref{evensplitpq0}. 
And so, $h$$-$1 copies of ${\cal S}_0$ exactly fill up the region~$R_1$ which is 
outside~$P$, inside~${\cal S}_0$ and between~$\rho_\ell$ and the line which 
supports~$e_2$. This splitting in exactly $h$$-$1 copies of ${\cal S}_0$ can be 
repeated for the region~$R_2$ defined by the continuations of~$e_2$ and~$e_3$
lying outside~$P$ and outside~$R_1$. We inductively define copies of the same
type of regions $R_i$ with $i\in \{1..p$$-$$3\}$. We have that $R_{i+1}$ is
outside~$P$ and outside all regions $R_j$ with $j\leq i$ and that it is defined
by the continuations of $e_{i-1}$ and~$e_i$. Now, the complement in ${\cal S}_0$
of~$P$ and all the $R_i$'s we have just defined is a new region~$S_1$. This region is
defined by~$e_{p-2}$, $e_{p-1}$ and~$e_p$, the angles between $e_{p-1}$ and its
neighbouring edges inside~${\cal S}_1$ begin both
$(h$$-$$1)\displaystyle{{4\hbox{\bf d}}\over q}$. 

Let $P_1$ be the reflection of~$P$ in~$e_{p-1}$. Denote by~$\rho_\ell$ the continuation
of $e_{p-2}$ and by~$\rho_r$ that of~$e_p$. The splitting of~${\cal S}_1$ is
a bit different from that of~${\cal S}_0$, but it relies on the same considerations.
Rename~$e_1$ the edge of~$P_1$ which is shared with~$P$ and denote by~$e_i$
the other sides of~$P_1$, $i\in\{1..p\}$, the numbering being increasing while
counter clockwise turning around~$P_1$. We notice that this time, in the
complement in~${\cal S}_1$ of~$P_1$, $\rho_\ell$ and $e_2$ define an angle
which is $(h$$-$$2)\displaystyle{{4\hbox{\bf d}}\over q}$, so that we split this
region, say~$R_1$ into~$(h$$-$$2)$ copies of~${\cal S}_0$. For the next regions
$R_i$, defined by~$e_{i}$ and~$e_{i+1}$ in the complement in~${\cal S}_1$ of~$P$
and the regions~$R_j$ for $j<i$,  $e_{i-1}$ and~$e_i$, but here 
for $i\in\{2..p$$-$$3\}$,
we have $h$$-$$1$ copies of~${\cal S}_0$. Now, the other ray~$\rho_r$ and~$e_p$
define in the complement of~$P$ in~${\cal S}_1$ another region~$R_{p-2}$ which
can also be split into~$(h$$-$$2)$ copies of~${\cal S}_0$. Now, what remains
in~${\cal S}_1$ after removing~$R_{p-2}$ is a region~$R_{p-1}$ which is a copy 
of~${\cal S}_1$.
Accordingly, we have split ${\cal S}_1$ into a copy of~$P$, $(p$$-$$2)(h$$-$$1)-2$
copies of~${\cal S}_0$ and one copy of~${\cal S}_1$, see~Figure~\ref{evensplitpq1}. 
}
\fi

\vskip 10pt
\setbox110=\hbox{\includegraphics[width=240pt]{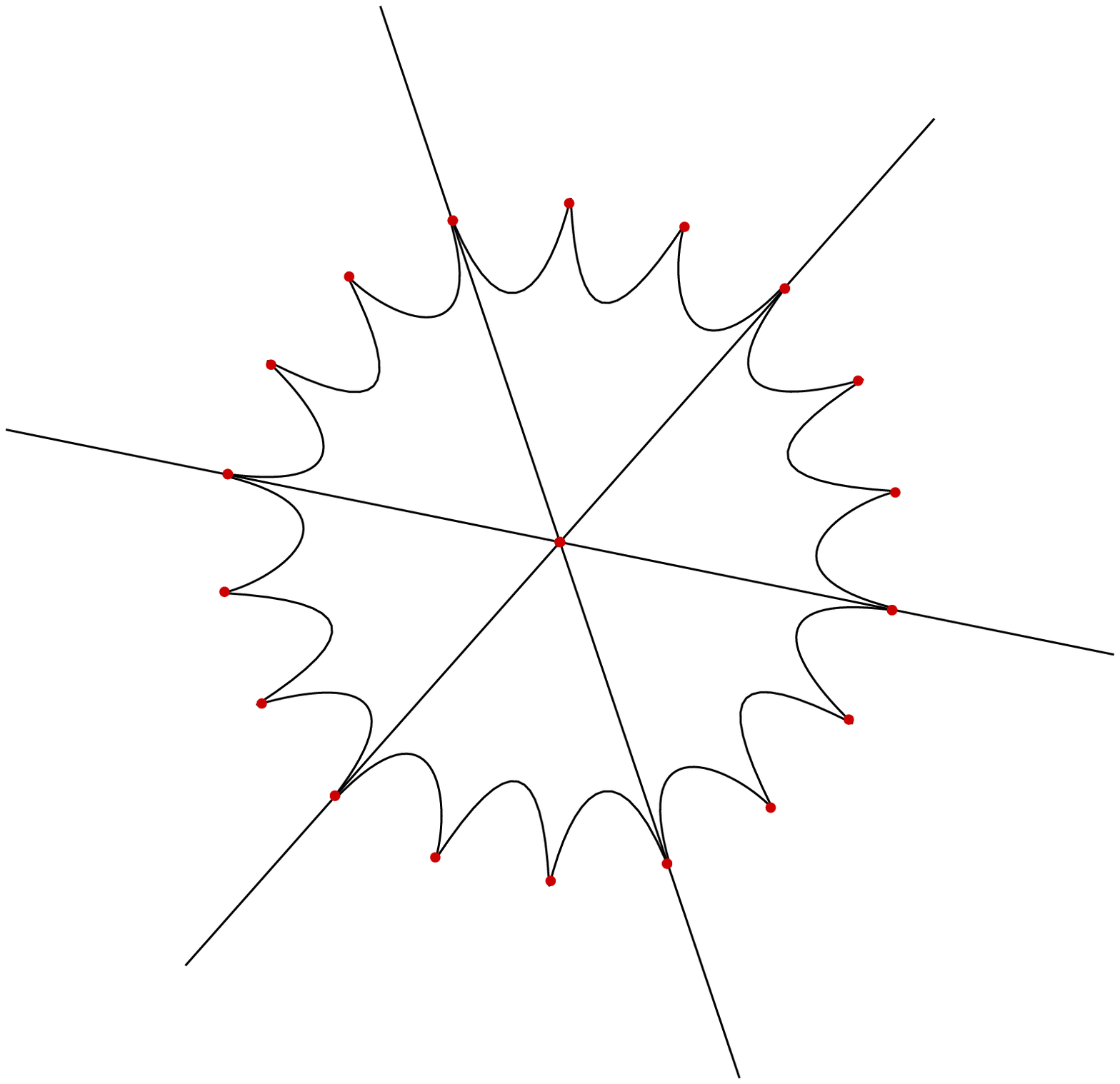}}
\vtop{
\ligne{\hfill
\PlacerEn {-290pt} {0pt} \box110
}
\vspace{-15pt}
\begin{fig} \label{evensplitpqeven}
\leurre
The even splitting of the tiling $\{p,q\}$ when $q$ is even.
\end{fig}
}
\vskip 10pt

  In~\cite{mmbook1} we indicate how the tiling $\{p,q\}$ can be split in order to 
prove that the tiling is combinatorial. From this we know that we can define
a tree which spans the restriction of the tiling to a copy of~${\cal S}_0$
in the conditions given in the above definition of such a sector.
We can see a representation of this splitting, which generalizes the notion of even
splitting to the case of the tiling~$\{p,q\}$ when $q$~is even, in 
Figure~\ref{evensplitpqeven}. 

   In~\cite{mmbook1} we also indicate the polynomial of the splitting, which is given
by \hbox{$X^2-((p$-$3)(h$$-$$1)$+$1)X-h+3$}, with $h$~defined by $q=2h$, from which 
we obtain that the
number~$u_n$ of nodes on the level~$n$ of the spanning tree satisfies the
following recurrence equation:
\hbox{$u_{n+2} = ((p$-$3)(h$$-$$1)$+$1)u_{n+1} - (h$+$3)u_n$}, with $u{-1}=0$
and $u_1=1$. From this, we define~$W$, the number of tiles observable in~${\cal S}_0$
as indicated in Sub-section~\ref{generalscheme}. 

   Now, for the evaluation of the number of tiles, note that the even splitting
gives us that exactly $q$~copies of~${\cal S}_0$ cover the hyperbolic plane with no
overlapping. As the number of nodes in the spanning tree is~$W$, we have 
that the number of tiles is $qW$. On another hand, the area of a tile
is here given by $(p$$-$$2).2\hbox{\bf d}-p\displaystyle{{4\hbox{\bf d}}\over q}$,
so that the total area which can be observed in these condition is
$2(pq$$-$$2(p$+$q))W\hbox{\bf d}$.

\vskip 10pt
\setbox110=\hbox{\includegraphics[width=240pt]{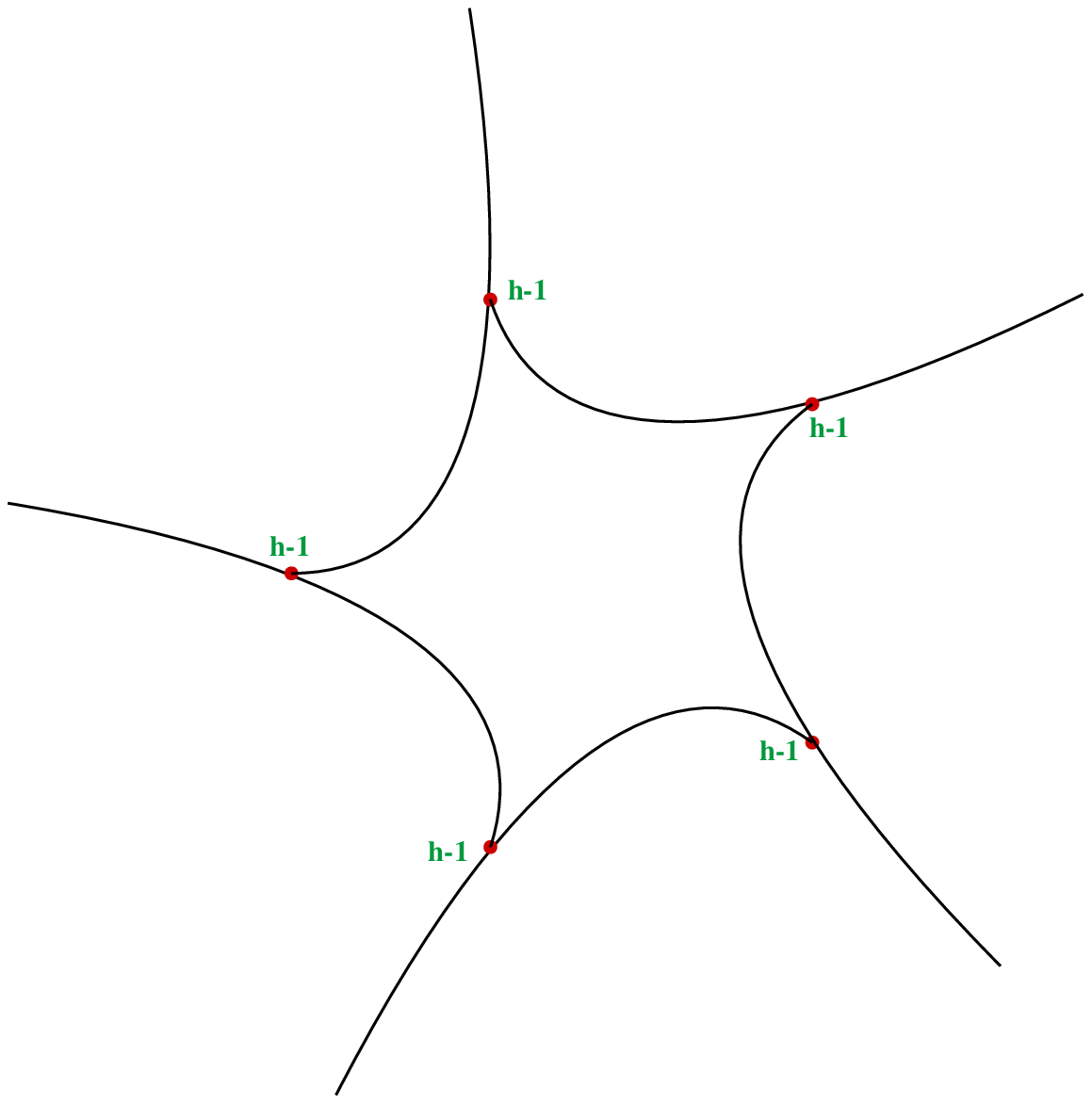}}
\vtop{
\vspace{-30pt}
\ligne{\hfill
\PlacerEn {-290pt} {0pt} \box110
}
\vspace{-45pt}
\begin{fig} \label{oddsplitpqeven}
\leurre
The odd splitting of the tiling $\{p,q\}$ when $q$ is even. Note that $h$$-$$1$ 
indicates
the number of copies of~${\cal S}_0$ which can be displayed in a fan rooted at the
nearby vertex, between the two rays defined by the sides meeting at the vertex, as 
indicated in the figure.
\end{fig}
}
\vskip 10pt

   Figure~\ref{oddsplitpqeven} indicates the splitting of the hyperbolic plane
organized around a central tile. We can see that the complement in the hyperbolic plane
of the central tile can be split into $p$~regions which are copies of each other.
Such a region is delimited by a vertex and two rays, each one supporting an edge of the
central tile abutting the vertex. The angle between the two rays is
$(h$$-$$1)\displaystyle{{4\hbox{\bf d}}\over q}$ and so, we can see that there is
there room for exactly $h$$-$1 copies of~${\cal S}_0$, all of them with the same vertex.
We say that such a region is a {\bf fan} of $h$$-$1 copies of~${\cal S}_0$. 
Accordingly, this time the number of tiles is $p(h$$-$$1)W+1$. We again
find the result we have found for the tilings $\{p,4\}$. Also note that the total
area which falls under observation is 
$(p(h$$-$$1)W+1)$
$((p$$-$$2).2\hbox{\bf d}-p\displaystyle{{4\hbox{\bf d}}\over q})$
= $2(h$$-$$1)\displaystyle{{pq\hbox{$-$}2(p\hbox{$+$}q)}\over q}W\hbox{\bf d}
+ 2\displaystyle{{pq\hbox{$-$}2(p\hbox{$+$}q)}\over q}\hbox{\bf d}
=(h$$-$$1)\displaystyle{s\over q}W\hbox{\bf d} + \displaystyle{s\over q}\hbox{\bf d}$,
where we have put $s=2(pq\hbox{$-$}2(p\hbox{$+$}q))$.

\subsubsection{The case when $q$ is odd: a first solution}
\label{generalodd1}

   In this case, we can no more use the mid-point lines which we considered
in the case when $q=3$. There is a solution which is thoroughly described
in~\cite{mmbook1}, but which is different, in its principle from what was done 
in the case when $q=3$ and also from the case when $q$~is even. There is a new 
solution in~\cite{mmarXiv} which we shall briefly describe in the next paragraphs.

\vskip 20pt
\setbox110=\hbox{\includegraphics[width=240pt]{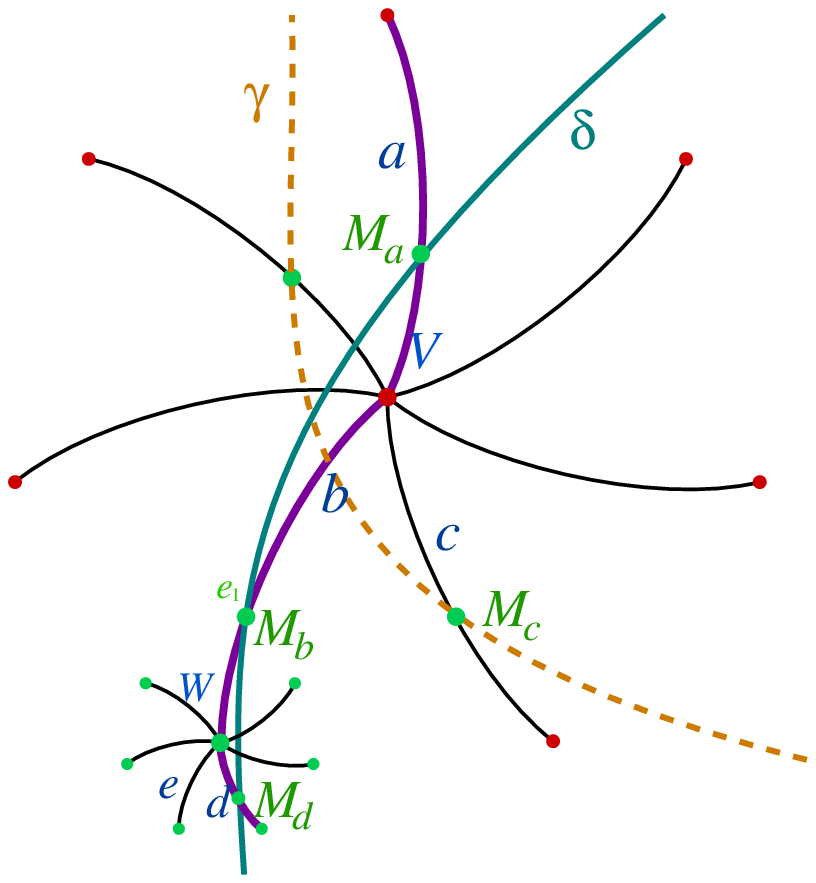}}
\vtop{
\vspace{-30pt}
\ligne{\hfill
\PlacerEn {-290pt} {0pt} \box110
}
\vspace{-25pt}
\begin{fig} \label{h_mid-point}
\leurre
The definition of the $h$-mid-point line.
\end{fig}
}
\vskip 10pt

   This solution consists in taking into consideration another mid-point line,
what is called a $h$-{\bf mid-point line} in~\cite{mmarXiv}, where
$h=\lfloor\displaystyle{q\over2}\rfloor$. This new line comes
from the following consideration. Around a vertex~$V$, we exactly have $q$~copies of
the polygon used for defining the prototile. Fix one edge~$a$ among the~$q$ ones 
abutting~$V$. There are exactly two edges~$b$ and~$c$ abutting~$V$ which makes an
angle $h\displaystyle{{2\pi}\over q}$ with~$a$. Let $M_a$, $M_b$ and~$M_c$ be the 
mid-points of~$a$, $b$ and~$c$ respectively, see Figure~\ref{h_mid-point}. 
Consider~$W$ the other end point of~$b$.
We can find two edges~$d$ and~$e$ abutting~$W$ and making with~$b$ the
same angle $h\displaystyle{{2\pi}\over q}$. Let~$d$ be the edge which is on the 
other side
of the line supporting~$b$ with respect to~$a$: $a$ is one half-plane defined by this
line while $b$ is in the other one. Let~$M_d$ be the mid-point of~$d$. It is easy to see
that the triangles $M_aVM_b$ and $M_bWM_d$ are equal and, consequently,
that the angles $(b,M_bM_a)$ and $(b,M_bM_d)$ are equal. As these angles are not on 
the same side of the line supporting~$b$, the points $M_a$, $M_b$ and~$M_d$ lie on 
the same line which is called an {\bf $h$-mid-point line}. Clearly, $M_a$ and~$M_c$ 
also define another $h$-mid-point line which is symmetric to this one under the 
reflection in the line~$\delta$ supporting~$a$. 

   This allows us to define what we shall call sector~${\cal S}_0$ when $q$~is odd.
Figure~\ref{evensplitpqodd} illustrates the display of $q$~copies of~${\cal S}_0$
around a central vertex. The figure also illustrates the construction we have just
defined in the previous paragraph. 

   We can see that here too, the hyperbolic plane can bee
split into $q$~sectors exactly.

\vskip 10pt
\setbox110=\hbox{\includegraphics[width=240pt]{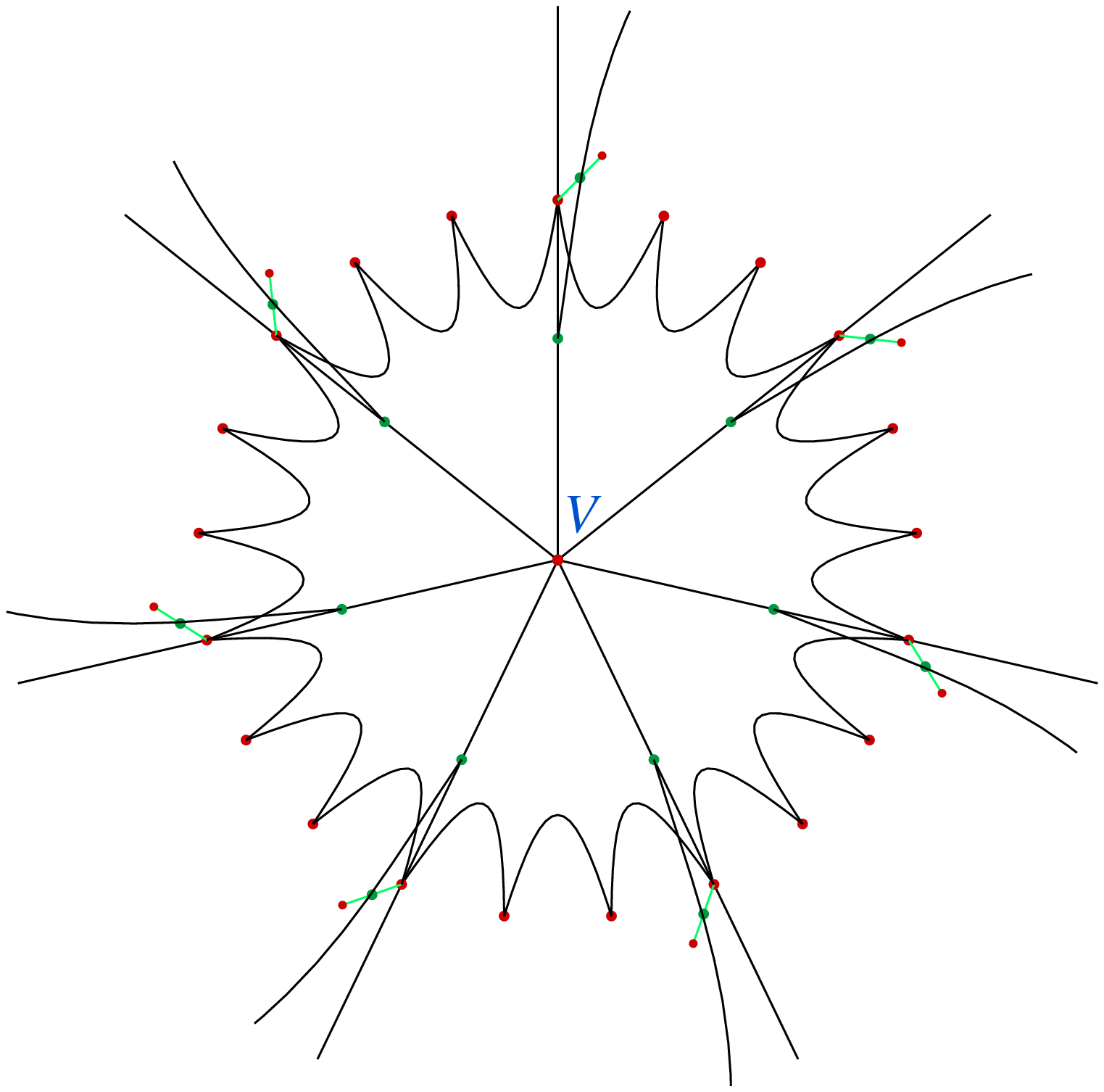}}
\vtop{
\vspace{-30pt}
\ligne{\hfill
\PlacerEn {-290pt} {0pt} \box110
}
\vspace{-25pt}
\begin{fig} \label{evensplitpqodd}
\leurre
The even splitting of the tiling $\{p,q\}$ when $q$ is odd. Note 
the $h$-mid-point lines which define the $q$~copies of~${\cal S}_0$ which 
share a common vertex. It goes from the mid-point of an edge abutting the central
vertex and through the mid-point of an edge making an angle 
$h\displaystyle{{2\pi}\over q}$ with the edge of the first mid-point we have just
considered.
\end{fig}
}
\vskip 10pt

   As shown in~\cite{mmarXiv}, it can be proved that we need three types of
region in order to obtain a combinatoric tiling, and we get that the
polynomial of the splitting is 
\hbox{$X^3 - ((p$$-$$3)(h$$-$$1)$+$1)X^2 - ((p$$-$$2)(h$$-$$1)$$-$$2)X - h$+3}.
This give rise to the following recurrence equation, which involves one more term
in the right-hand side:
\hbox{
$u_{n+3} = ((p$$-$$3)(h$$-$$1)$+$1)u_{n+2} + ((p$$-$$2)(h$$-$$1)$$-$$2)u_{n+1}
- (h$$-$$3)u_n$}, where $u_n$ is the number of nodes which are on the level~$n$
of the spanning tree. This allows us to define~$W$, the number of observable
tiles in the sector~${\cal S}_0$.

Consequently, we have $qW$ tiles which can be observed in this way. Now, the 
computation of the area of the basic polygon is the same as in the case 
when $q$~is even so that we find
the same expression for the total area under observation, namely
$sW${\bf d}, with again $s= 2(pq$$-$$2(p$+$q))$. 

   Let us now consider the odd splitting for the tiling $\{p,q\}$ when $q$~is odd.

\vskip 10pt
\setbox110=\hbox{\includegraphics[width=240pt]{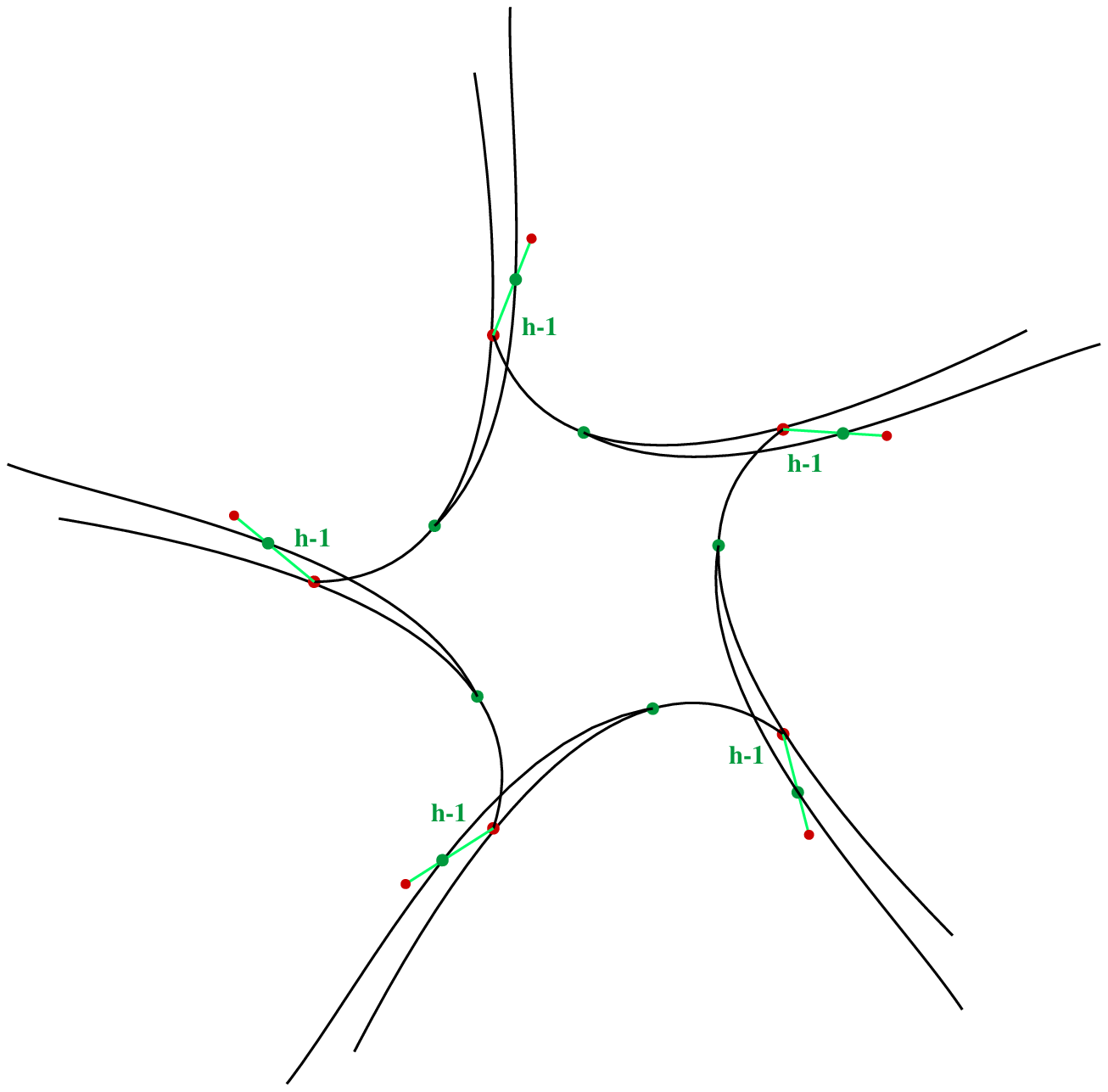}}
\vtop{
\vspace{-30pt}
\ligne{\hfill
\PlacerEn {-290pt} {0pt} \box110
}
\vspace{-25pt}
\begin{fig} \label{oddsplitpqodd}
\leurre
The odd splitting of the tiling $\{p,q\}$ when $q$ is odd. Note 
the $h$-mid-point lines which define the $q$~fans of $h$$-$$1$~copies of~${\cal S}_0$ 
which are displayed around a central tile. 
\end{fig}
}
\vskip 10pt

   Figure~\ref{oddsplitpqodd} illustrates the display of $p$~fans of $h$$-$1~copies
of~${\cal S}_0$ around a central tile. We note that this 
splitting is different from the one we considered when $q=3$. The definition
of the sectors under the general definition would define a region which
is the union of two sectors in terms of those of the heptagrid. We shall soon go back 
to this point.

   Accordingly, we can see that now we have $p(h$$-$$1)W+1$ tiles under observation.
Accordingly, the total area which is observable is defined by the same formula 
for the odd splitting as in the case when $q$~is even.

   We can now formulate the results:

\begin{thm}\label{thpqodd1}
The number of tiles of the tiling $\{p,q\}$, when $q\geq 4$ is
$qW$ in the even splitting, independently of the parity of~$q$, where
$W=U_\mu$, and the total observable area is
$sW\hbox{\bf d}$, where $s=2(pq-2(p$+$q))$.
With the odd splitting, the number of tiles which can be observed is
$p(h$$-$$1)W+1$, where $h$~is defined by 
$h=\lfloor\displaystyle{q\over2}\rfloor$, corresponding to the possible 
observation of a total area which is 
$p(h$$-$$1)\displaystyle{s\over q}W\hbox{\bf d}+\displaystyle{s\over q}\hbox{\bf d}$.
\end{thm}  

   As we can see from the areas indicated in the theorem, the odd splitting allows us
to observe a much bigger area:  $(h$$-$$1)\displaystyle{s\over q}$ is a bit smaller
than $\displaystyle{1\over 2}$, but closer to this value as $q$~increases. Besides,
$p$~is at least~3, so that $(h$$-$$1)\displaystyle{p\over q} > 1$, from which we 
conclude that 
$p(h$$-$$1)\displaystyle{s\over q}W\hbox{\bf d}+\displaystyle{s\over q}\hbox{\bf d}
> sW\hbox{\bf d}$.
\subsubsection{The case when $q$ is odd: another solution}
\label{generalodd2}

   In~\cite{mmarXiv}, we indicate another splitting of the tiling $\{p,q\}$
in the case when $q$~is odd. This gives an alternative proof that the tiling is
combinatoric. The advantage is that the number of basic regions is now two
instead of three in thee previous solution, so that the polynomial of the splitting
has degree~2. In fact, the new splitting comes from the fact that one region
of the previous solution can easily be split into the two others. And this region
is the generalization to the general case $\{p,q\}$, with $q$~odd, of what was
done in the heptagrid. If we go back to Figure~\ref{h_mid-point}, the dashed 
line~$\gamma$ is a rotated image of the line~$\delta$, and a sector is defined in
this way in the previous solution. In the solution indicated in this sub-subsection
and which is detailed in~\cite{mmarXiv}, the sector is defined by the lines~$M_aM_b$
and $M_aM_c$. This means that the second line is not a rotated image of the first
one around the vertex~$V$ but the reflection of the first line in the bisector of the
angle at~$V$ between the two sides of the polygon meeting at~$V$. Note that
this is the generalization of the definition of a sector given for the heptagrid.
The second region is ${\cal S}_1$ as in the previous solution.

\vskip 10pt
\vskip 10pt
\setbox110=\hbox{\includegraphics[width=240pt]{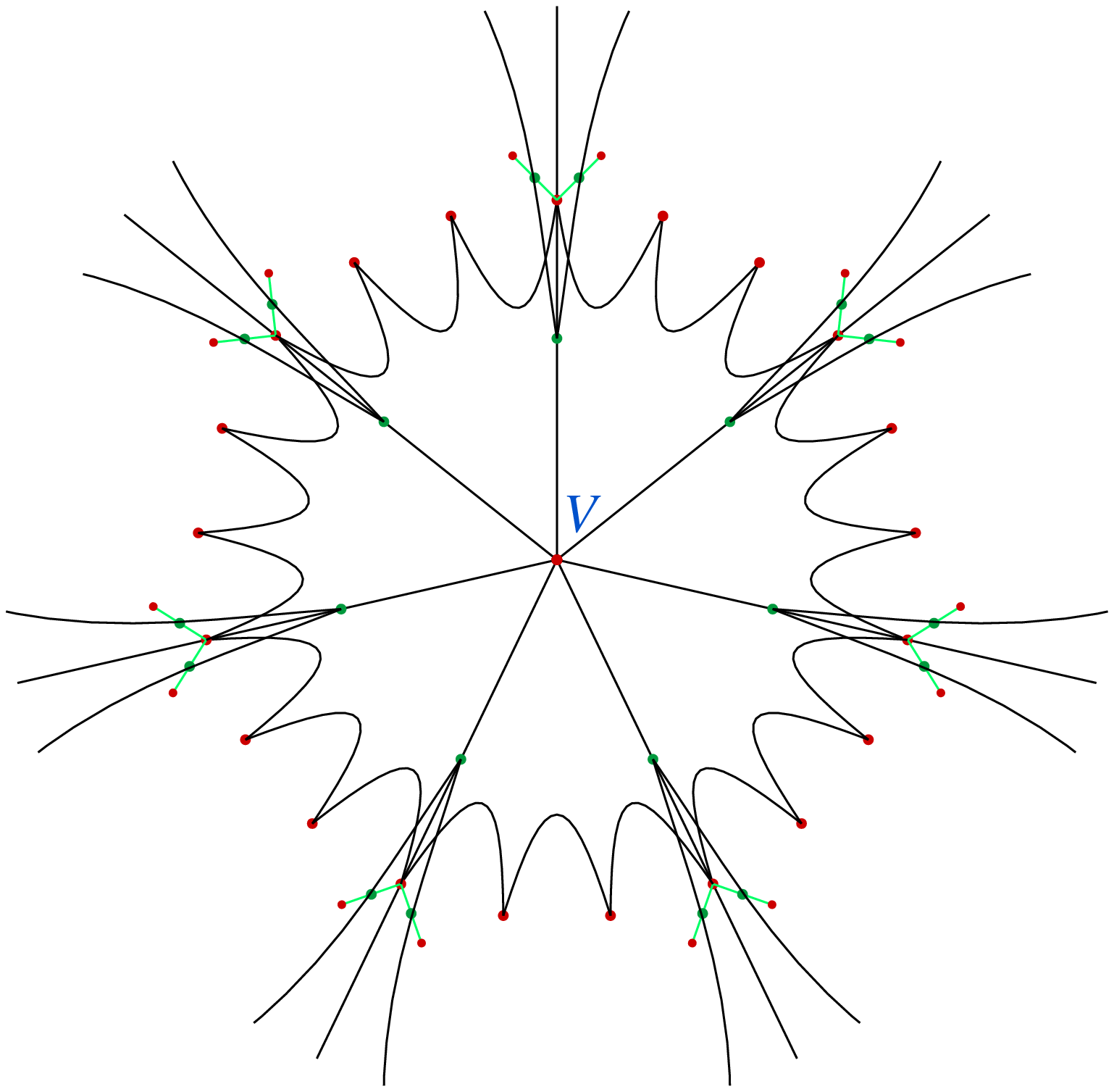}}
\vtop{
\vspace{-30pt}
\ligne{\hfill
\PlacerEn {-290pt} {0pt} \box110
}
\vspace{-25pt}
\begin{fig} \label{evensplitpqoddv2}
\leurre
The even splitting of the tiling $\{p,q\}$ when $q$ is odd, alternative version. 
Note the $h$-mid-point lines which define the $q$~copies of~${\cal S}'_0$ which 
share a common vertex. The union of the $q$~copies of~${\cal S}i'_0$ around~$V$
do not cover the plane.
\end{fig}
}
\vskip 10pt
   Now, following the computations of~\cite{mmarXiv}, the polynomial of the splitting 
is: \hbox{$X^2 - ((p$$-$$3)(q$$-$$3)$+$1)X - q$+7}, which is also very different
from what we obtain in~\cite{mmbook1}. From the polynomial, we get the recurrent 
equation, again with two terms on the right-hand side:
\hbox{$u_{n+2} = ((p$$-$$3)(q$$-$$3)$+$1)u_{n+1} + (q$$-$$7)u_n$}.
Again, we define~$W$ as indicate in Sub-section~\ref{generalscheme}, the
basic region being now ${\cal S}'_0$.

   This time, we can see that there is a difference between the even and the odd
splittings. Indeed, if we consider $q$~copies of~${\cal S}'_0$ around~$V$, they
do not cover the hyperbolic plane. As we can see in Figure~\ref{evensplitpqoddv2},
in between two contiguous copies of~${\cal S}'_0$, there is room for another
copy of~${\cal S}'_0$ which is a proper part of~${\cal S}'_0$ It is easy
to see that the height of the tree spanning~${\cal S}'_0$ being $\eta$,
the eight of the tree spanning this other copy of~${\cal S}'_0$ is $\eta$$-$1. 
Let us set $W_1=U_{\eta-1}$ and let us denote by~${\cal S}^1_0$ the set of
tiles spanned the tree of height~$\eta$$-$1. Then, in the even splitting, 
the plane is split into $q$~copies of~${\cal S}'_0$ and $q$~copies
of~${\cal S}^1_0$ which cover exactly the plane with no overlapping,
the number of tiles being $q(W$+$W_1)$. For the odd splitting, we have the same 
$p$~copies of a fan of $h$$-$1 copies of~${\cal S}_0$ and $h$$-$1 of~${\cal S}^1_0$, 
which means $2(h$$-$$1)$ copies of~${\cal S}'_0$
as ${\cal S}_0$ can exactly be split into two copies of ${\cal S}'_0$.
Accordingly we have $p(h$$-$$1)(W$+$W_1)$+1 tiles. 
Now, the area of the regular polygon is 
$2(pq$$-$$2(p$+$q))\displaystyle{\hbox{\bf d}\over q}$ in both cases. This allows us 
to state the following result:
 
\begin{thm}\label{thpqodd2}
The number of tiles of the tiling $\{p,q\}$ in the alternative splitting, 
when $q\geq 5$ and $q$~is odd, is $q(W$$+$$W_1)$~under the even splitting 
and $p(h$$-$$1)(W$$+$$W_1)+1$ under the 
odd splitting, where $W_1=U_{\eta-1}$ while $W=U_\eta$. 
The observable total area is given by
$s(W$$+$$W_1)${\bf d} in the even splitting and by
$p(h$$-$$1)\displaystyle{s\over q}(W$$+$$W_1)\hbox{\bf d} 
+ \displaystyle{s\over q}\hbox{\bf d}$
in the odd splitting. 
\end{thm}

   As in the previous solution, we can notice that the odd splitting allows us to 
observe a bigger area than in the even one.

\section{{\Large Conclusion}}
\label{conclusion}

   It is interesting to see the difference between the even and odd splitting 
in the study of the number of tiles which can be observed using the infinite
numeral system. In all cases, we can 
count more than $\grossone$ tiles. 
Also, in all cases, the odd splitting gives more tiles and, accordingly a greater
area. This is particularly striking when $q\geq 5$ in the tilings $\{p,q\}$.

While looking at the families $\{p,4\}$ and $\{p$+$2,3\}$ which have the same spanning
tree, it is worth noticing that, despite the fact that in the
odd splitting, the tiling $\{p$+$2,3\}$ gives access to more tiles than the
tiling $\{p,4\}$, the total observable area is bigger with the tiling~$\{p,4\}$: this
corresponds to the fact that the basic polygon is significantly bigger in this tiling.  
Another remarkable point is the fact that the ratio between the different areas we 
can observe is the same, independently of~$p$.

   Now, when $q$~is odd, the observable total area
is bigger in the second solution than in the previous one: it is twice the previous
one for the even splitting and almost twice too for the odd splitting. This indicates
the interest of the second solution which is also useful for its possibility
to simulate action at a distance: indeed, in this splitting, each node of the
tree has sons which correspond to a tile which is not in contact with the tile of
the father, even by a vertex only.

It appears that the infinite
numeral system based on \grossone{} gives a precise tool
to measure properties which allow us to introduce a distinction in various splittings
of the same tiling which, classically, are all equivalent.
It would be interesting to explore the possibilities given by this system 
for other criteria.


\section*{Acknowledgment}

I wish to express special thanks to Yaroslav {\sc Sergeyev} for his great attention
to this work.  
 
\bibliographystyle{eptcs}

\end{document}